\documentclass{aa}  

\usepackage{graphicx}
\usepackage{txfonts}
\usepackage{lipsum}
\usepackage{subcaption}         
\usepackage{lscape}             
\usepackage{placeins}           

\begin{document}

   \title{Diagnosing the solar atmosphere through the Mg I b$_2$ 5173 \AA\ line}

   \subtitle{I. Nonlocal thermodynamic equilibrium inversions versus traditional inferences}

   \author{A. L. Siu-Tapia\inst{1,2}
        \and L. R. Bellot Rubio\inst{1,2}
         \and D. Orozco Su\'arez\inst{1,2}
        }

   \institute{Instituto de Astrof\'isica de Andaluc\'ia (IAA-CSIC), Apdo. 3004, 18080 Granada, Spain\\
             \email{azaymisiu@gmail.com}
            \and Spanish Space Solar Physics Consortium, Spain\\ }

   \date{Received November 29, 2024}

 
  \abstract
   {}
   {
   We examined the capabilities of methods based on the weak-field approximation and line bisectors
to extract fast and reliable information about the height stratification of the magnetic field and line-of-sight velocities, respectively, from high spatial resolution observations of the Mg I b$_2$ line at 5173 {\AA}. } 
   {The Mg I b$_2$ line
was analyzed alongside the Fe I 6173 \AA\ line to help constrain the physical conditions of the photosphere. Additionally, we present the first high-resolution 
inversions of the Mg I b$_2$ line under nonlocal thermodynamic equilibrium (NLTE) conditions conducted over
a large field of view using a full-Stokes multiline approach. 
To determine the optimal inversion strategy, we performed several tests on the Mg I b$_2$ line using the Fourier Transform Spectrometer atlas profile before applying it to our observations.}
   {The good correlations between the traditional methods and the NLTE inversions indicate that the
weak-field approximation is generally a reliable diagnostic tool at moderate field strengths for the
rapid inference of the longitudinal magnetic field from the Mg I b$_2$ line. 
In contrast, line bisectors exhibit poorer correlations with the NLTE inferred plasma velocities, suggesting that they might not be suitable for deriving velocity gradients from the Mg I b$_2$ line.
Furthermore, to
accurately derive the thermodynamic properties of the solar atmosphere from this line, the
more complex, and time-consuming, NLTE Stokes inversions are necessary. This work also
provides observational evidence of the existence of low-lying canopies expanding above bright magnetic
structures and pores near the low chromosphere.}
   {}


   \maketitle

\section{Introduction}
\nolinenumbers
 
  Unlike the photosphere, the solar chromosphere offers a very limited selection of spectral lines with sufficient magnetic sensitivity. This makes the inference of chromospheric magnetic fields a very challenging task. 
 In the mid-chromosphere, magnetic fields are commonly probed using polarimetric measurements of the Ca II 8542 \AA\ line \citep[e.g.,][]{ delaCruz2017, Gosic2018, Esteban2019, Quintero2019, Siu2020} despite its weak polarization signals.
  However, the situation is even more difficult in the so-called temperature minimum region, which extends from the upper photosphere to the low chromosphere, where very few spectral lines form \citep[][]{Plaskett1931, Waddell1963, Stenflo1984, Stenflo2000}. Among these few lines, for example, Mg I b$_2$ 5173 {\AA}, KI D$_1$ 7698 {\AA}, Na I D$_1$ 5896 \AA,\ and D$_2$ 5890 {\AA}, 
  those that could potentially provide additional information have not yet been fully explored, leaving the magnetism at these heights largely unknown. This gap limits our understanding of the magnetic field topology and its connectivity between the photosphere and the chromosphere, which is crucial for understanding the energy transfer and interactions between the two layers.

 A good candidate for probing the magnetic field in this region is the Mg I b$_2$ line at 5173 {\AA}, which is magnetically sensitive and forms across a wide range of heights \citep[e.g.,][]{Stenflo1984, Lites1988, Stenflo1997, Quintero2018}, effectively bridging the gap  between the photospheric lines and the chromospheric Ca II 8542 \AA\ line.
 With the ability to observe multiple spectral lines simultaneously,  studies at these heights are already  possible with unprecedented spatial and temporal resolution thanks to instruments in solar observatories such as 
 the \textit{Daniel K. Inouye }Solar Telescope \citep[DKIST;][]{Rimmele2019} and SUNRISE III \citep[][]{Barthol2011}, and in the future with the European Solar Telescope \citep[EST;][]{Quintero2022}.

\begin{table}[ht!]
\caption{Description of the CRISP/SST observations.}        
\label{tab:1}      
\centering                                      
\begin{tabular}{c c c c c c c c }          
\hline\hline 
Atom &    $\lambda_0$  &   $g_{\mathrm{eff}}$ &   $N_{\lambda}$ &  $\Delta\lambda$   & $\Delta\lambda_{\mathrm{i,f}}$   &  $\Delta\lambda_\mathrm{C}$    & $N_{\mathrm{acc}}$   \\
 &    ({\AA}) &    &    &   (m{\AA})  &  (m{\AA}) &     (m{\AA}) &    \\

\hline
\hline

Fe I & 6173.33 & 2.5 & 18 & 28 &  $\pm$224 &  $+532$  & 18 \\
Mg I  & 5172.68 & 1.75 & 14 & 50\tablefootmark{*}  &  $\pm$500 &  $-700$  & 27 \\
Ca II & 8542.09 & 1.1 & 18 & 100  &   $\pm800$ &  $+2400$  & 4 \\
H$\alpha$ & 6562.81 & 1.05 & 21 & 100  &  $\pm$900 &  $\pm1000$ &   4 \\

\hline                                   
\hline     
\end{tabular}
\tablefoot{ From left to right, the columns indicate the atomic species of the four observed spectral lines, the central wavelength ($\lambda_0$), the
effective Land\'e factor ($g_{\mathrm{eff}}$), the number of wavelength points within the line ($N_{\lambda}$), the step size 
($\Delta\lambda$), the initial and final wavelengths within the spectral line ($\Delta \lambda_{\mathrm{i,f}}$), the continuum wavelength point relative to the line center ($\Delta\lambda_\mathrm{C}$), and the number of accumulations per wavelength ($N_{\mathrm{acc}}$).
\\
\tablefoottext{*}{The spectral window at $\pm$100 m\AA\ from the line core was sampled with $\Delta\lambda=$50 m{\AA}, while $\Delta\lambda=$100 m{\AA} was used outside the line core, up to $\pm$500 m{\AA}.}}
\end{table}

 To date,  very few polarimetric observations of the Mg I b$_2$ line have been reported in the literature \citep[e.g.,][]{Lites1988, Choudhary2007, Martinez2009, Deng2010, Gosic2018, Morosin2020}. The core of this spectral line forms in the low chromosphere under nonlocal thermodynamic equilibrium (NLTE) conditions \citep[e.g.,][]{Quintero2018}, which may partly explain why it has not been extensively used for magnetic field inferences.
 Properly interpreting the chromospheric radiation requires solving the radiative transfer problem under a NLTE scheme, that is to say, taking the atomic states and the populations of the different energy levels into account, which leads to a nonlocal and nonlinear problem \citep[e.g.,][]{delaCruz2017}.
 Currently,  
 several inversion codes are capable of addressing the radiative transfer problem under NLTE assumptions \citep[e.g.,][]{Socas2015, Milic2018, delaCruz2019, Ruiz2022}.  
 However, these codes require substantial computational resources and time, particularly when inverting large datasets, such as time series or snapshots covering large fields of view (FOVs).

 An alternative method for estimating the magnetic field configuration in those cases is the weak field approximation \citep[WFA;][]{Deglinnocenti1973}. The WFA can be applied when the Zeeman splitting is significantly smaller than the Doppler width of the spectral line, a condition often satisfied by chromospheric lines since they tend to be broad. However, this approximation also assumes the absence of gradients in the velocity or magnetic field along the line of sight (LOS), which may not hold true everywhere in the chromosphere. Nevertheless, it has been proven that, if applied to narrow spectral windows within chromospheric lines, the WFA can provide insight into the  stratification of the longitudinal magnetic field in the height formation range \citep[e.g.,][]{Centeno2018, Siu2020}.
 Similarly, line bisectors \citep[LBs;][]{Maltby1964} can serve as a rapid alternative tool for inferring the height stratification of LOS plasma velocities in certain scenarios \citep[e.g.,][]{Schlichenmaier2004, Bellot2006, Esteban2016, Gonzalez2020}.
 
 In this work we
evaluated the performance of different methods --- namely, the WFA, LBs, LTE and NLTE inversions --- for the inference of magnetic field and LOS velocity gradients in different magnetic structures, using full-Stokes observations of the Mg I b$_2$ line at high spatial resolution.  
 We confront the results obtained from the different methods  and assess the validity of each of them in the different magnetic regimes.
 The inferred magnetic field gradients also allow us to study the low-lying magnetic canopies that expand above the magnetic structures.
 
 This paper is organized as follows: In Sect. 2 we describe the observations and the post-processing of the data. In Sect. 3 we explain the various methods employed to infer the physical parameters from our spectropolarimetric observations and describe a series of NLTE inversion tests that helped us determine the configuration setup that best works with the characteristics of our observations. In Sect. 4 we present the main results and compare the different methods. Finally, in Sect. 5 we discuss our findings and provide our conclusions.

\section{Observations} \label{sec:style}

The observations  
were carried out on July 22, 2019, at the Swedish 1 meter Solar Telescope \citep[SST;][]{Scharmer2003} from 07:20:04 UT to 07:27:10 UT, using the CRisp Imaging Spectrometer \citep[CRISP;][]{Scharmer2008}. 
We observed a bipolar region at an heliocentric angle of $\sim 26^{\circ}$ (56$\arcsec$ S, 426$\arcsec$ W), over a FOV of nearly $55\arcsec\times60\arcsec$ with a plate scale of $0.057\arcsec$ per pixel.

 \begin{figure*}
   \centering
  \includegraphics[width=0.95\hsize]{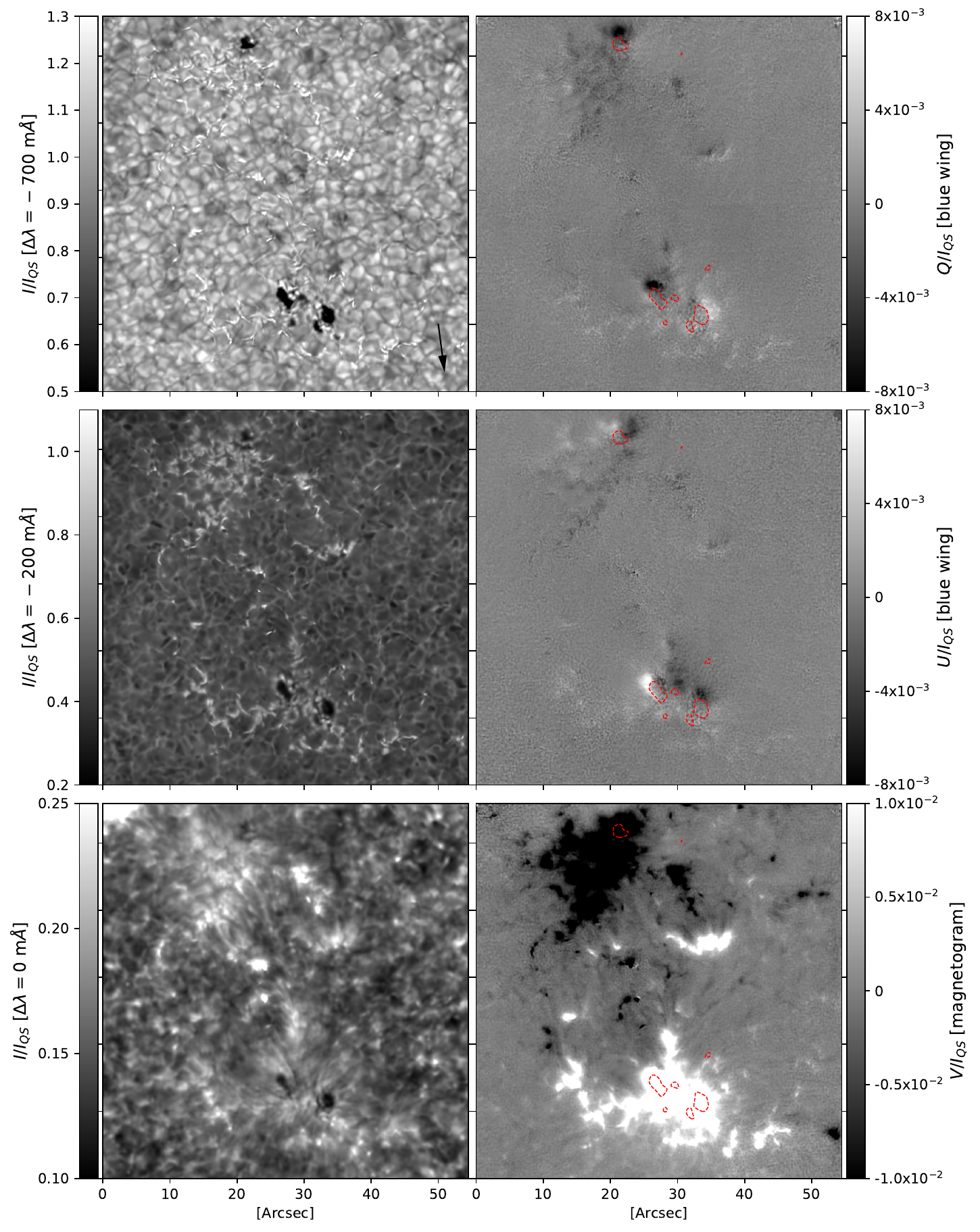} 
      \caption{Intensity and polarization maps normalized to the continuum intensity of the quiet Sun, $I_{QS}$. They show a bipolar region observed with CRISP in the Mg I b$_2$ 5173 \AA\ line at 07:20 UT. Left: Intensity maps at different wavelengths. Right: Maps of  $Q$ and $U$ averaged over three wavelengths in the blue wing ($\Delta\lambda=-100$, $-50$, and $0$ m{\AA}), and a Stokes $V$ magnetogram at $\pm50$ m\AA\ from the core. The black arrow points toward the disk center. The red contours in the right panels delineate the boundaries of the pores.
              }
         \label{fig:1}
   \end{figure*}

 \begin{figure}[t]
   \centering
    \includegraphics[width=0.95\hsize]{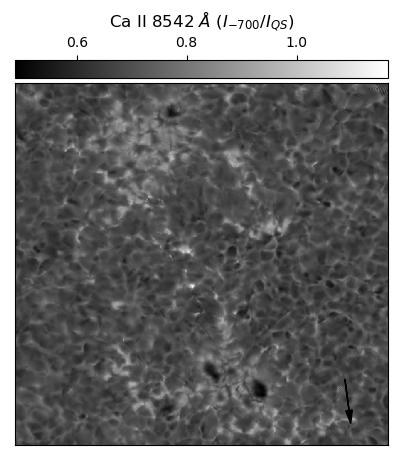}
      \caption{Intensity map at -700 m\AA\ from the core of  the Ca II 8542 \AA\ line. The observation was taken at 07:20 UT.
              }
         \label{fig:1b}
   \end{figure}

The dataset consists of four spectral scans of the photospheric and chromospheric lines indicated in Table \ref{tab:1},
 with a temporal cadence of $\sim$ 116 
s between consecutive scans. 
In particular, the Mg I b$_2$ line was sampled with a denser grid in the $\pm$100 m\AA\ region around its core compared to its wings in order to extract more detailed information from the low chromosphere and temperature minimum region of the atmosphere. Also, while  the Fe I and Mg I b$_2$ lines were both observed in polarimetric mode, the Ca II and H$\alpha$ lines were recorded in spectroscopic mode.

 The data were reduced using the 
SSTRED  
pipeline \citep[][]{Lofdhal2021}
and the multi-object multi-frame blind deconvolution  \citep[MOMFBD;][]{Vannoort2005} technique for image restoration.
We focus here on the analysis of the Fe I 6173 \AA\ and the Mg I b$_2$ 5173 \AA\ lines, which were  observed in polarimetric mode.
For these lines we corrected the residual crosstalk from $I$ to $Q$, $U$, and $V$ by using the Stokes parameters recorded 
in their corresponding continuum wavelength points, which are indicated in Table \ref{tab:1} (for simplicity, we refer to the first recorded wavelength of the Mg I b$_2$ line as the continuum wavelength,  although the point at $\Delta \lambda=-700$ m\AA\ actually  lies in the far blue wing, still within the line).

The left column of Fig. \ref{fig:1} shows intensity maps at different wavelengths within the Mg I b$_2$ line portraying the different layers in the atmosphere to which this line is sensitive. 
The top left panel shows the normalized continuum intensity and provides a typical photospheric view over the full FOV, which comprises a bipolar active region formed by a few small pores  surrounded by bright structures or plage regions. The FOV also covers a portion of the quiet Sun granulation 
containing various isolated bright points.
The middle left panel displays the map at $\Delta\lambda=-200$ m\AA\ and shows the reversed granulation pattern characteristic of the mid photosphere at heights $\sim$130-140 km \citep[e.g.,][]{Rodriguez1999, Leenaarts2006, Cheung2007}. 
Interestingly, the intensity map at this wavelength looks remarkably similar to that at  -700 m\AA\ from the core of the Ca II 8542 \AA\ line, which is shown in Fig. \ref{fig:1b}.
Such a similarity indicates that the formation height range of these two lines overlap partially, and therefore complement each other, which is of great importance for studying the transition between the photosphere and the chromosphere. \citet[][]{Rutten2011} also present a comparison of Mg I b$_2$, Na I D$_1$, and Ca II 8542 \AA\ and show the spectral images from each line overlap at different wavelength positions \citep[see also, e.g.,][]{Morosin2020}.
The bottom left panel of Fig.\ \ref{fig:1} corresponds to the line-core intensity map, which displays a more chromospheric configuration, with canopy-like structures that expand outward from the pores. 

The linear and circular polarization maps 
depicted in the right column of Fig.\ \ref{fig:1}  show the largest Mg I b$_2$ Stokes $Q$, $U$, and $V$ signals within the active region (i.e., in the pores and their surrounding bright regions).
However,  the largest linear polarization signals are found in the periphery of the pores rather than in their central cores. In contrast, in the quiet Sun granulation both the linear and circular polarization signals are at the noise level of these CRISP observations, except in localized areas that can be identified as isolated bright points from the intensity maps.

To reduce the noise level of the observed polarization signals, we convolved the data with a $3\times3$ low-pass filter kernel so that each pixel in the resulting images has a value equal to the average value of its neighboring pixels in the original image. The noise level in the final images are $\sigma_Q=0.0017$, $\sigma_U=0.0018$, and $\sigma_V=0.0013$ for the Mg I b$_2$ line, and $\sigma_Q=0.0022$, $\sigma_U=0.0019$, and $\sigma_V=0.0021$ for the Fe I line.

\section{Methods}
In this section we describe the various methods used to infer the physical parameters from our SST observations. First, we explain the traditional methods, namely, the inversions of the Fe I line under LTE conditions, the WFA applied to different spectral windows within the Mg I b$_2$ line, and the estimation of LOS velocities from the Mg I b$_2$ LBs and line-core Doppler shifts. Afterward, we describe different NLTE inversion tests carried out on the Mg I b$_2$ line using the Fourier Transform Spectrometer \citep[FTS;][]{Neckel1984} atlas profile to help determine the inversion strategy that best fits our SST observations. Finally, we describe the two-line NLTE inversions performed over the entire FOV.

\subsection{LTE inversions of the Fe I 6173 \AA\ line}
We performed a simple one-component inversion of the Fe I 6173 \AA\ line  to get information of the photospheric magnetic field and LOS plasma velocities. To that end, we employed the SIR inversion code \citep{Ruiz1992}, which solves the radiative transfer equation (RTE) under the assumption of LTE to infer the temperature ($T$), LOS velocity ($v_{LOS}$), magnetic field strength ($B$), inclination ($\gamma$), and azimuth ($\phi$) from the observations. 

The inversions are run over five cycles considering a maximum of five nodes in temperature, and only one node in all the other physical parameters. The initial model atmosphere uses a FALC 
 temperature stratification \citep{Fontela1993} that extends from $\log(\tau)=1$ to $-4$, and starts with constant values of $v_{LOS}$ (1 km s$^{-1}$), $B$ (100 G), $\gamma$ (35$^{\circ}$), and $\phi$ (35$^{\circ}$).

%
%

\begin{table}
\caption{Atomic parameters of the three strongest blends in the Mg I b$_2$ 5173 \AA\ line. }        
\label{tab:2}      
\centering                                      
\begin{tabular}{c c c c c c}          
\hline\hline                        
Atom & $\lambda_0$ ({\AA}) & $\log$ $(gf)$ & $L_1$ & $U_1$ & $g_{\mathrm{eff}}$\\

\hline                                   
\hline                                             
Fe I & 5171.59 & -1.793 & $^{3}\mathrm{F}_{4}$ &  $^{3}\mathrm{F}_{4}$ & 1.25 \\
Fe I & 5171.67 & -1.912 & $^{3}\mathrm{D}_{3}$ &  $^{1}\mathrm{G}_{4}$ & 0.50 \\
Ti I & 5173.74 & -1.118 & $^{3}\mathrm{F}_{2}$ &  $^{3}\mathrm{F}_{2}$ & 0.67 \\
\hline                                   
\hline     
\end{tabular}
\tablefoot{From left to right, the columns indicate the corresponding atomic species, the central wavelength, the  $\log(gf)$ of the transition, and the spectroscopic notation of the lower and the upper levels. Data taken from the database of the 
\citet{Nist2024}
and verified for consistency in \citet{Kurucz1995}. The last column contains the effective Land\'e factor computed assuming L-S 
coupling by \citet{Quintero2018}.}
\end{table}

\subsection{Weak field approximation for the Mg I 5173 \AA\ line}

The WFA is valid when the thermal broadening in a given spectral line is much larger than its Zeeman splitting \citep[e.g.,][]{Deglinnocenti2004}. In that situation, one can apply a perturbative scheme to the RTE to simplify the problem, so the first-order perturbation yields the following solution for the longitudinal magnetic field:

\begin{equation}
B_{LOS}=\frac{\sum_k \frac{\partial I(\lambda_k)}{\partial\lambda_k} V(\lambda_k)}{\alpha\sum_k(\frac{\partial I(\lambda_k)}{\partial\lambda_k})^2},
\label{eq:8}
\end{equation}

\noindent where $\alpha=-4.67\times10^{-13}g_{\mathrm{eff}}\lambda_0^2$, with $g_{\mathrm{eff}}$ being the effective Land\'e factor and $\lambda_0$ the central wavelength of the line in {\AA}. The index $k$ spans over the different wavelength points sampled within the spectral line. 

The second-order perturbation leads to expressions that relate the linear polarization with the first and second derivatives of the line intensity through the transverse component of the magnetic field, $B_T$, which can be estimated as follows:

\begin{equation}
B_{T}=\left(\frac{\sum_k \frac{4}{3} \frac{L(\lambda_k)}{C_T} \left|\frac{1}{\lambda_k-\lambda_0}\right| \left|\frac{\partial I(\lambda_k)}{\partial\lambda_k}\right| }{\sum_k \left|\frac{1}{\lambda_k-\lambda_0}\right|^2 \left|\frac{\partial I(\lambda_k)}{\partial \lambda_k}\right|^2}\right)^{1/2}  \text{for }  \lambda_k \neq \lambda_0,
\label{eq:8T}
\end{equation}

\noindent with $L(\lambda_k)=\sqrt{Q^2(\lambda_k)+U^2(\lambda_k)}$ and $C_T=(4.6686 \times 10^{-10} \lambda_0^2)^2 G_{\mathrm{eff}}$ is a constant that depends on the spectral line, where $G_{\mathrm{eff}}$ is the Land\'e factor for the transverse magnetic field \citep{Centeno2018}.

We used the WFA to estimate the height variation of the longitudinal field and of the magnetic field inclination by applying Eqs. \ref{eq:8} and \ref{eq:8T} to three different spectral windows within the Mg I $b_2$ line: $\pm500$ m{\AA}, $\pm300$ m{\AA}, and $\pm100$ m{\AA}. The narrower the spectral window around the line center, the higher the atmospheric layer it probes.

\subsection{Line bisector method}
Since spectral lines get contributions from different layers of the solar atmosphere, large gradients with height in the LOS plasma velocities can produce strong asymmetries in the intensity profile. Those velocity gradients can be somehow quantified by measuring the so-called LB \citep[e.g.,][]{Maltby1964}.
The LB at a specific intensity level, $I_{bis}$, is given by the central wavelength position of the line, $\lambda_{bis}$,  at such an intensity level. 

We calculated the bisectors of the Mg I b$_2$ line at eight 
 different intensity levels by using a linear interpolation of the observed profiles.  
The intensity difference between the line core and the local continuum corresponds to 100$\%$.
The selected intensity levels were placed between 10$\%$ and 80$\%$ of the spectral line depth, in steps of 10$\%$.
For higher intensity levels the LBs  become unreliable due to a larger noise sensitivity near the continuum level \citep{Schlichenmaier2004}.
 The LB Doppler shifts were then used to compute the $v_{LOS}$ associated with the different intensity levels. 

The deepest bisector levels sample the plasma motions at the highest layers the line is sensitive to. However, exactly at the line core, the LB method cannot be applied. Instead, we apply a Bezier fit \citep[e.g.,][]{bezier1968, farin2002} to the profile in the line-core wavelengths, so that the analytic minimum of the fitted function provides us with a good estimate of the line core wavelength position, $\lambda_{lc}$. We use these values as a proxy for the Doppler velocities induced by plasma motions along the LOS at different atmospheric heights as follows:

\begin{equation}
v_{LOS}=\frac{c \Delta\lambda_{D}}{\lambda_0},
\end{equation}

\noindent where $c$ is the speed of light, $\lambda_0$=5172.6843 {\AA}, and $\lambda_D$ stands for either the $\lambda_{bis}$ or $\lambda_{lc}$.
The reference wavelength for zero velocity is taken from  the average intensity profile over entire dataset\footnote{To our knowledge, there are no studies in the literature on how the Mg I b$_2$ line is impacted by solar convective blueshift \citep[e.g.,][]{Lohner2019} in its core wavelengths.  However, this effect is expected to be small as the core of this line forms in the low chromosphere, where convective motions are significantly weaker than in the photosphere. The FTS atlas profile 
of the Mg I b$_2$ line shows a redshift in the line core position relative to its laboratory rest wavelength, corresponding to $\sim$446 m s$^{-1}$. 
 Theoretical predictions and observations indicate that the solar gravitational redshift, as observed from Earth, is approximately 636 m s$^{-1}$ \citep[e.g.,][]{LoPresto1991, GonzalezHernandez2020}. Thus, the convective blueshift in the Mg I b$_2$ line is likely less than 200 m s$^{-1}$. 
In this study, we neglected the effects of convective blueshift and use the average position of the Mg I b$_2$ line core as the zero-velocity reference for estimating LOS velocities from 
this spectral line.}.

\begin{table*}
\caption{Configuration of the different NLTE inversion tests.  }        
\label{tab:3}      
\centering                                      
\begin{tabular}{c c c c c c c}          
\hline\hline                        
Test &Spectral lines  & Spectral window  & Wavelength sampling   & Mg atomic model & OF & NLTE threshold\\
 &    &  ({m\AA}) & (m{\AA})  &  &  & \\

\hline                                   
\hline                                             
1 & Mg 5173, blends  & $\pm$ 3000 & 10 &  6 level & yes &  10,8,5,2,1,0.1,0.01,0  \\
2  & Mg 5173, blends & $\pm$ 3000 & 10 & 6 level & yes & 0  \\
3 & Mg 5173, blends  & $\pm$ 3000 & 10 & 6 level & no & 0  \\
4 & Mg 5173, blends & $\pm$ 3000 & 10   & 13 level & yes & 0 \\
5 & Mg 5173  & $\pm$ 3000 & 10 & 13 level  & yes & 0  \\
6 &  Mg 5173, Fe 6173  & $\pm$ 3000, $\pm$ 1000 & 10 & 13 level  & yes & 0  \\
7 & Mg 5173, Fe 6173  & $\pm$ 800, $\pm$ 560 & 10 & 13 level  & yes & 0  \\
8 & Mg 5173, Fe 6173  & $\pm$ 800, $\pm$ 560 & 50, 28 & 13 level & yes & 0  \\
\hline                                   
\hline     
\end{tabular}
\tablefoot{The columns, from left to right, indicate the inversion test number,  the spectral lines included in the inversions (where "blends" refer to those in the Mg I b$_2$ line listed in Table \ref{tab:2}; smaller or unknown blends in the Mg line were masked in all tests), the spectral window of the inverted FTS atlas profiles, the wavelength sampling of the FTS profiles, the atomic model used for the Mg atom, the inclusion of opacity fudge (OF), and the NLTE threshold used in each of the eight inversion cycles. In test 8, the spectral window and wavelength sampling refer to those of the inversion grid only, since the FTS profiles were inverted using only the wavelength points that coincide with those of our SST observations as detailed in Table \ref{tab:1}, i.e., using  
14 wavelength points in the Mg line and 
18 in the Fe line; see Fig.\ \ref{fig:1c}.}
\end{table*}

 \begin{figure*}
   \centering
    \includegraphics[width=0.95\hsize]{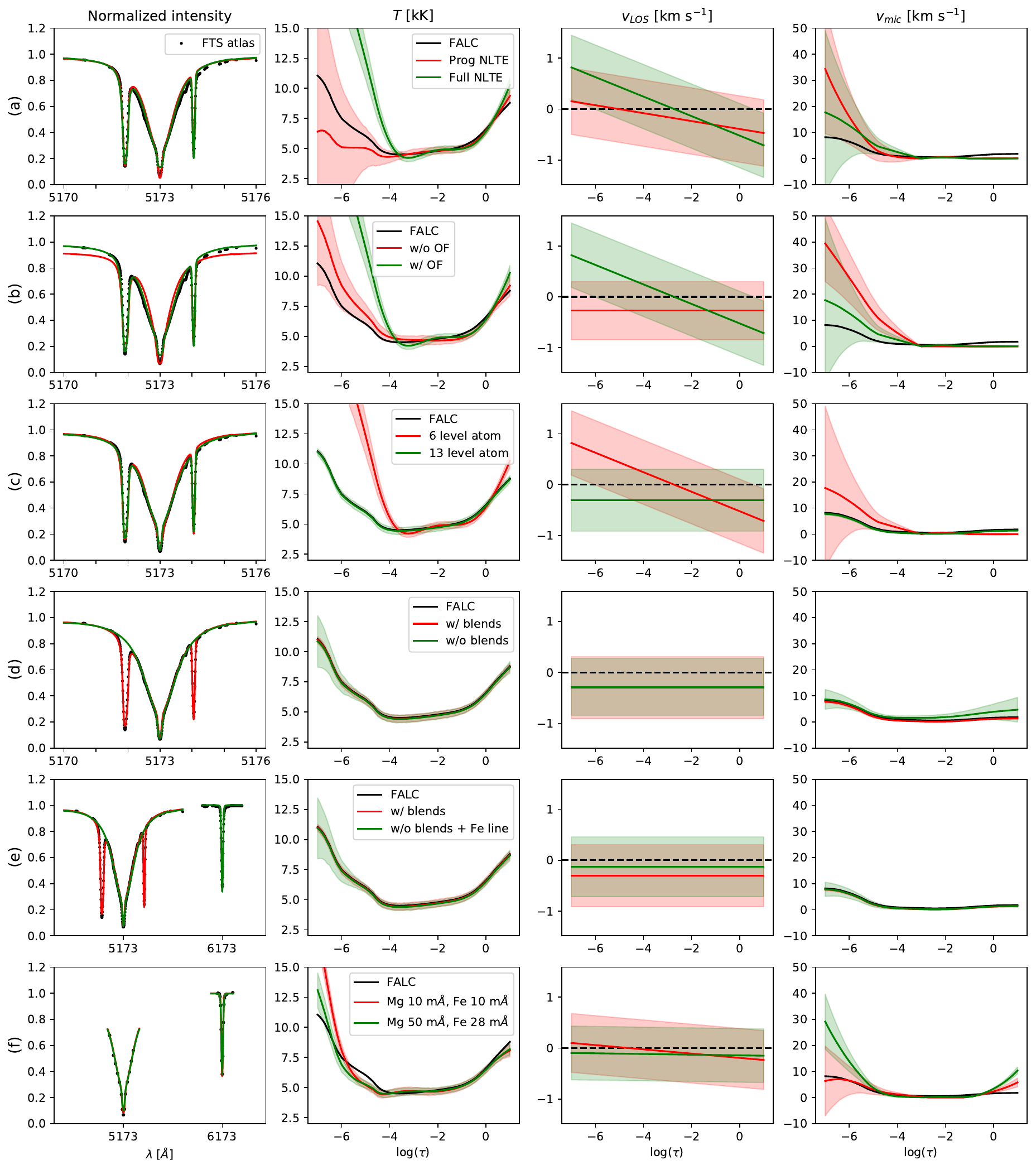} 
      \caption{NLTE inversions of the FTS atlas profiles using different setups. The FTS profiles are shown with black dots and have a wavelength sampling of 10 m\AA\ for both the Mg I 5173  \AA\ and Fe I 6173 \AA\ lines, except for one of the inversions in the last row, as specified in the legends. 
      The best-fit profiles and the resulting atmospheric stratification from the different inversions are presented with colored solid lines, and the shaded areas indicate the estimated uncertainties. Each row compares the following setups: (a) a progressive NLTE inversion (test 1, red lines) versus a full NLTE inversion (test 2, green); (b) the effects of including OF (test 3, red) versus not using OF (test 2, green); (c) an inversion using the 6-level Mg atomic model (test 2, red) versus an inversion using the 13-level Mg atomic model (test 4, green); (d) the inclusion of some Mg blends (test 4, red) versus their exclusion (test 5, green); (e) the inclusion of Mg blends (test 4, red) versus their exclusion but  including the photospheric Fe 6173 \AA\ line instead (test 6, green); and finally, the effect of wavelength sampling in a smaller spectral window of $\pm 800$ m\AA\ in Mg and $\pm560$ m\AA\ in Fe, comparing a spectral grid of 10 m\AA\ in both lines (test 7, red) to a  grid similar to our SST observations with a step of 50 m\AA\ for Mg and 28 m\AA\ for Fe  (test 8, green). For the detailed configuration used in each of these inversion tests, refer to Table \ref{tab:3}.
       }
         \label{fig:1c}
   \end{figure*}

\subsection{NLTE inversions: Tests with FTS}

We used the DeSIRe code  \citep[Departure coefficient Stokes Inversion based on Response functions;][]{Ruiz2022} 
to perform NLTE inversions of both the Mg I 5173 \AA\ line (treated in NLTE) and the Fe I 6173 \AA\ line (treated in LTE) simultaneously.
The DeSIRe code, which combines the RH forward synthesis code \citep{Uitenbroek2001} and the SIR inversion code \citep[][]{Ruiz1992}, enables the inversion of spectral lines formed under NLTE conditions, such as the Mg I b$_2$ line. It is currently the fastest available NLTE inversion code.

This study marks the first instance of high-resolution spectropolarimetric observations of the Mg I b$_2$ line being inverted using a NLTE code. To determine the best strategy for these inversions, we used the FTS atlas intensity profiles --- which represent the quiet Sun (i.e., no magnetic fields) --- to test different initial setups and study the effects of single-line and multiline inversions, the inclusion of blends and photospheric lines, the impact of spectral resolution, the performance of different atomic models, and the effects of opacity fudge \citep[OF; ][]{Bruls1992} .

All tests were initialized with a FALC model atmosphere \citep[][]{Fontela1993}, using parameters of $v_{LOS}=1$ km s$^{-1}$, $B=100$ G, $\gamma=35^{\circ}$, and $\phi=35^{\circ}$. The inversions were run over eight cycles, inverting only the temperature (from three to six nodes), the microturbulent velocity (from one to five nodes), and the LOS velocity (from one to five nodes). We first inverted the Mg I b$_2$ intensity profile alone, using a spectral window covering $\pm$3 \AA\ around the line core position. This window includes three strong photospheric blends — two from Fe I in the blue wing and one from Ti I in the red wing (with atomic parameters listed in Table \ref{tab:2}); smaller or unknown blends were masked. The wavelength step size was set to 10 m{\AA}.

To estimate the errors in the retrieved parameters, which DeSIRe does not provide, we iteratively introduced small random variations in the input model parameters and repeated each inversion 100 times. The standard deviation of the resulting atmospheres is taken as an estimate of the error for each physical parameter, and is shown as colored shaded regions in the figures.

Figure \ref{fig:1c} presents some results from the different NLTE inversion tests specified in Table \ref{tab:3}. 
The first row of Fig.\ \ref{fig:1c}  shows the best fits and the resulting atmospheric stratifications obtained using two different approaches: a progressive NLTE inversion and a full NLTE inversions (test 1 and test 2 in Table  \ref{tab:3}). In test 1, the first cycle assumes LTE, but in subsequent cycles, the NLTE threshold\footnote{DeSIRe  allows the user to specify a NLTE threshold value, which can range from 0 to 10. This value indicates the minimal percentage of the temperature perturbations used to reevaluate the departure coefficients and synthesize the line with RH between cycles. If the NLTE threshold is set to 10, the code performs the inversion in LTE. Conversely, a NLTE threshold of 0 means the code performs a full NLTE inversion  \citep{Ruiz2022}.} decreases progressively, with only the final cycle performed in full NLTE (red lines). In contrast,  test 2 applies a complete NLTE scheme in all 8 cycles (green lines). Both approaches use a simplified atomic model with just 6 levels and 3 line transitions \citep{Quintero2018} and employ OF 
 to more accurately model the UV radiative over-ionization in the Mg I b$_2$ line forming region. 
  Although the best fit from test 1 is better in the line core than  test 2, the resulting temperature profile from test 1 is unrealistic, as it remains almost flat in the upper atmospheric layers 
and shows no clear chromospheric temperature rise.
 We consider that inversions from test 2 provide a more realistic atmospheric stratification, while still acceptably fitting most of the atlas intensity profile, but note that the two solutions are compatible within the uncertainties up to about $\log(\tau)\sim-4$. 
 Therefore, for subsequent inversions, we adopted a full NLTE scheme (NLTE threshold = 0) in all of the cycles.

The second row of Fig.\ \ref{fig:1c} shows the results of inversions performed with and without OF (tests 2 and 3 in Table \ref{tab:3}, represented by green and red lines in the figure, respectively). Except for the line core, the fit is noticeably better at nearly all wavelengths when OF is included. This result reinforces the necessity of OF for properly inverting the Mg I b$_2$ line, especially in the wings.
The reason is that most magnesium in the solar photosphere and chromosphere is singly ionized, and the ionization of Mg is highly dependent on the UV radiation field. Unfortunately, modeling this field is challenging due to the large number of lines at UV wavelengths. 
Together, they significantly increase the opacity in the UV.
This increased opacity brings the source function closer to the Planck function and reduces over-ionization, which is caused by a suprathermal UV/extreme-UV radiation field.
The OF 
simulates the effect
 of these numerous lines by artificially enhancing the UV opacity through a multiplicative factor. Without OF, the ionizing radiation field becomes too strong, leading to an underestimation of the Mg I population and resulting in a line that is too narrow due to insufficient opacity (H. Uitenbroek, private communication, April 2022).
Therefore, all subsequent inversions are performed with the application of OF.

The third row in Fig.\ \ref{fig:1c} displays the results from inversions based on different atomic models. The first one uses the simplified 6-level atom (test 2, red lines), which was employed in all previous tests and is included as one of the default models for the Mg atom in the DeSIRe distribution. The second one uses a more complex atomic model with 13 levels and 44 transitions,  presented and described in detail by \citet[test 4, green lines]{Quintero2018}. Our results show that the 13-level atom provides a significantly better fit for both the line core and the wings compared to the 6-level atom. Additionally, the  retrieved atmosphere is simpler and closely resembles the initial FALC stratifications. This model also appears to require fewer nodes in $T$, $v_{LOS}$, and $v_{mic}$  to achieve a good  fit to the intensity profile. 
It is not surprising that NLTE inversions perform better with a more complete atomic model, as \citet{Quintero2018} demonstrated that the departure coefficients computed for the simplified 6-level atom result in a NLTE underpopulated ground and transition levels compared to the more complex 13-level atom. The departure coefficients presented in that work not only highlight the importance of NLTE effects for the levels involved in the Mg I b$_2$ line transition but also demonstrate that the 13-level atom provides more accurate NLTE results.
For these reasons, we performed the subsequent inversions using the 13-level atom despite the computation time being considerably longer --- approximately seven times slower than with the 6-level atom. This increased computation time has a noticeable impact, especially when performing NLTE inversions for a large number of profiles and,  in particular, when those inversions consider the full Stokes vector.

The forth row of Fig.\ \ref{fig:1c} displays the results of inversions that include and exclude the photospheric blends in the wings of the Mg I b$_2$ line (test 4 and test 5, red and green lines, respectively). In both cases, the quality of the fits is comparable, and the resulting atmospheric stratifications are remarkably similar. The main difference between the two is observed in the photospheric values of  $v_{mic}$, which are slightly larger when the blends are excluded from the inversions. Additionally, the associated uncertainties in  $v_{mic}$ are also larger in this case.  This discrepancy is likely due to the lack of photospheric information to better constrain the parameter values in the lower layers of the photosphere when the blends are not taken into account. This highlights the importance of including photospheric information during the inversions. However, the spectral range and sampling of the Mg I b$_2$ line extracted from the FTS atlas seems to be appropriate to constrain the photospheric temperatures and velocities without the need to include any photospheric blend in the fit.

The fifth row of Fig.\ \ref{fig:1c} compares the inversion results when the Mg I b$_2$ line blends are included (test 4, red lines) with the case where these blends are excluded, but instead, the Fe I 6173 Å line is inverted simultaneously (test 6, green lines). 
Interestingly, the quality of the fits is comparable in both cases, and the retrieved atmospheric stratifications are largely consistent with each other and with the FALC model. Some discrepancies arise in the uppermost layers around $\log(\tau)=-6$, but neither of these lines is sensitive to those optical depths. 
Therefore, both approaches — either including photospheric blends or incorporating a photospheric line like Fe I 6173 {\AA\ }— provide useful information from the lower layers of the atmosphere, helping constrain the values of the physical parameters at those heights.
In real observations, blends are not always present, so caution is needed when interpreting the inversion results for $v_{mic}$ and possibly the temperature in the lower layers unless a photospheric line is included in the inversion process.

Finally, the last row of Fig.\ \ref{fig:1c} shows the results of the two-line inversion for which we used a reduced spectral window for both lines to make them resemble our SST observations (i.e., $\pm$ 800 m\AA\ around the Mg I b$_2$ line core and $\pm$ 560 m\AA\ around the Fe I line core). 
The first test used a spectral sampling of 10 m\AA\ (test 7, red lines), and the second one employed the exact same sampling as in our observations, as indicated in Table \ref{tab:1} (test 8, green lines). A better quality fit is obtained with the denser spectral sampling, particularly in the Mg I b$_2$ line core, emphasizing the importance of having sufficient spectral resolution at these wavelengths. Nevertheless, the retrieved atmospheric stratifications are remarkably similar below $\log(\tau)\sim -5$ in both cases and  remain consistent with the FALC model.

 \begin{figure*}[hbtp!]
   \centering
   \includegraphics[width=0.95\hsize]{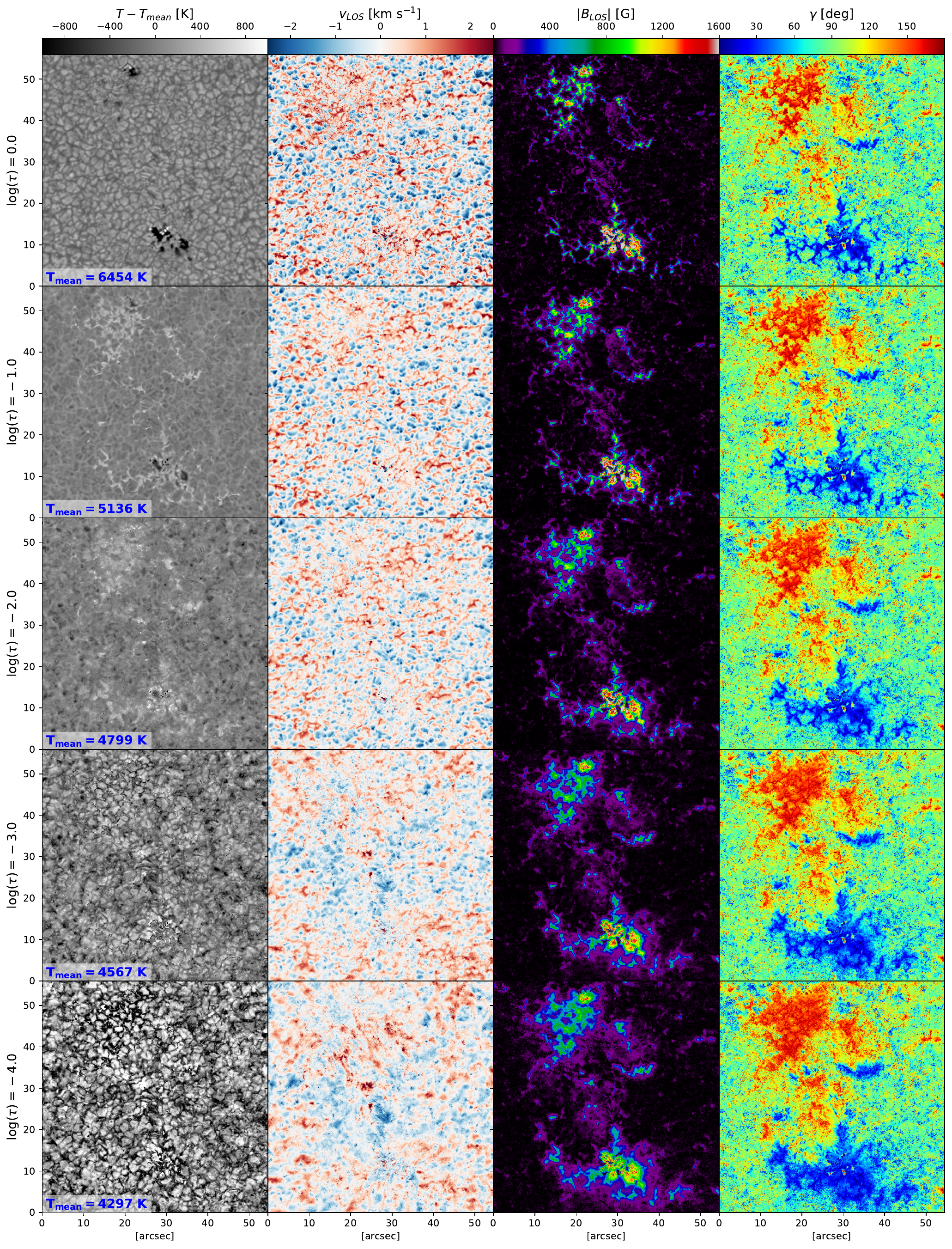}
      \caption{Results of the two-line NLTE inversion at five selected optical depths. From top to bottom, the maps correspond to $\log(\tau)$=0, -1, -2, -3, and -4. From left to right: Temperature variation [K], LOS velocity [km s$^{-1}$], unsigned longitudinal magnetic field [G], and magnetic field inclination [deg]. The mean temperature values of the FOV, $T_{mean}$, are indicated in the bottom left corners of the first-column panels for the corresponding optical depths. }
         \label{fig:NLTE_FOV}
   \end{figure*}

 \begin{figure}[t]
   \centering
    \includegraphics[width=0.95\hsize]{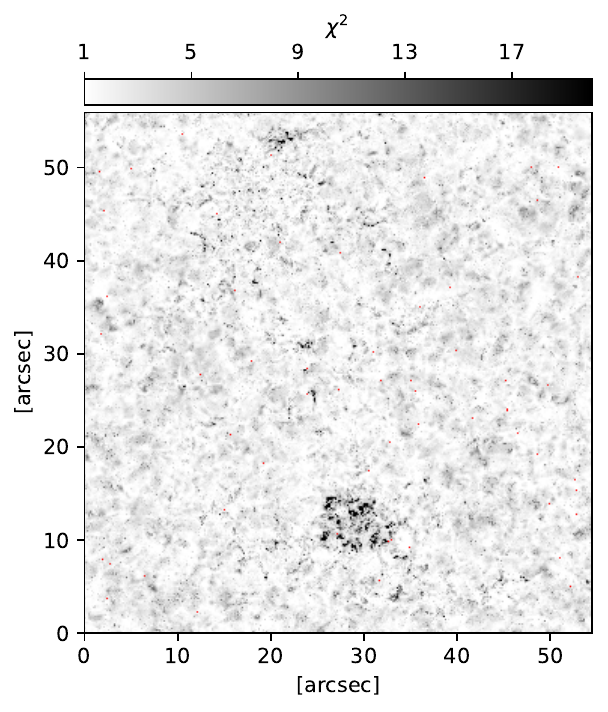} 
      \caption{Map of the merit function, $\chi^2$, resulting from the two-lines NLTE inversions. The red pixels in the map indicate those places where the inversion did not converge to a good solution ($\sim0.06\%$ of the FOV).
              }
         \label{fig:chi2}
   \end{figure}

 \begin{figure*}[hbtp!]
   \centering
  \includegraphics[width=\hsize]{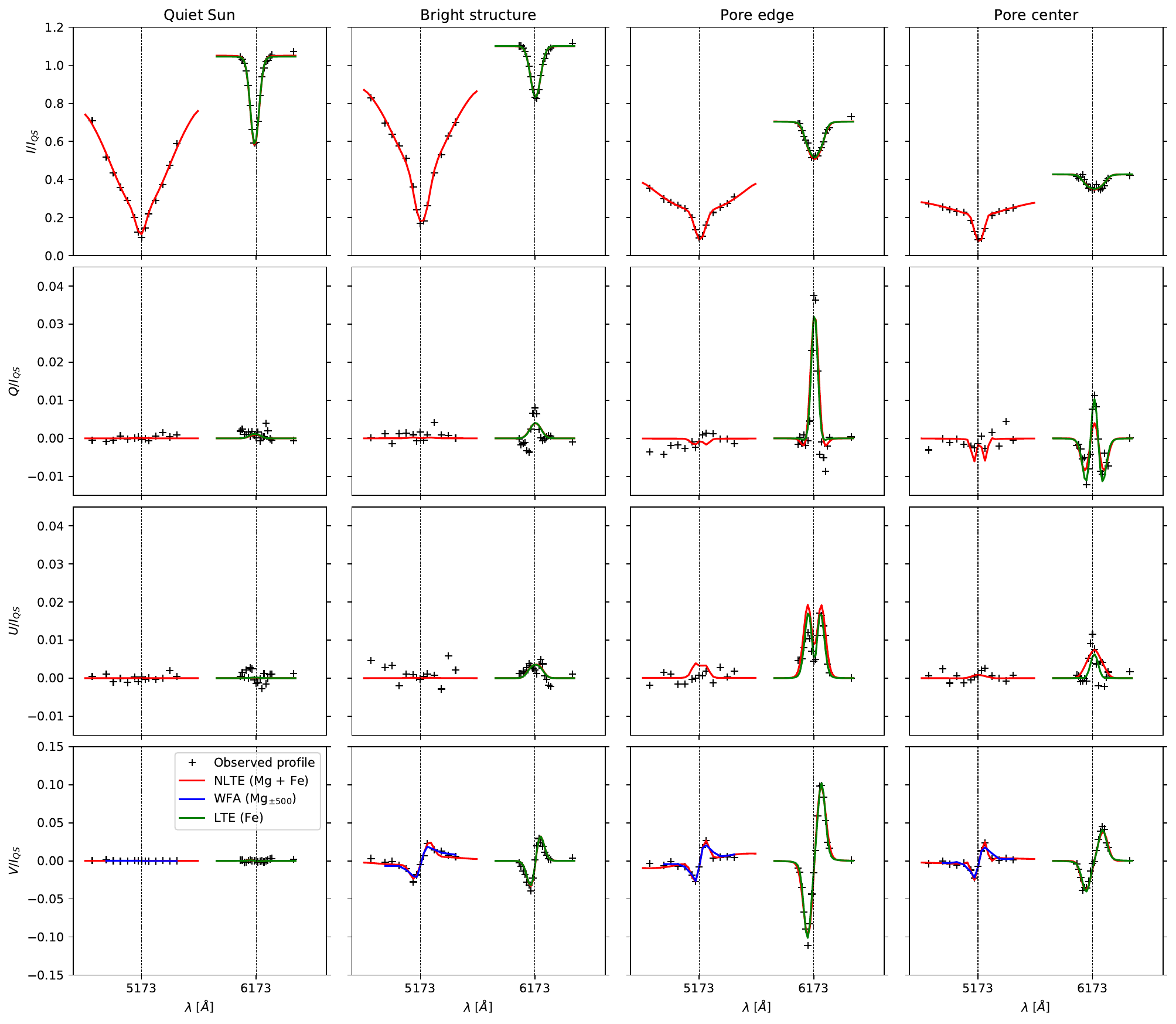} 
      \caption{Examples of Stokes profiles from different regions in the FOV (black markers) and best fits obtained with the different methods: two-line NLTE inversions (red lines), WFA using the spectral region $\pm500$ m\AA\ around the Mg line core (blue), and LTE inversion of the Fe line (green). Panels display, from top to bottom, the Stokes $I$, $Q$, $U$, and $V$ profiles normalized to the average continuum intensity in the quiet Sun, $I_{QS}$. From left to right, each set of Stokes profiles corresponds to a pixel in the quiet Sun, in a bright magnetic structure, in a pore, and in a pore's center. The resulting atmospheric stratifications are shown in Fig. \ref{fig:atms_SP}. 
              }
         \label{fig:fits_SP}
   \end{figure*}

 \begin{figure*}[hbtp!]
   \centering
     \includegraphics[width=\hsize]{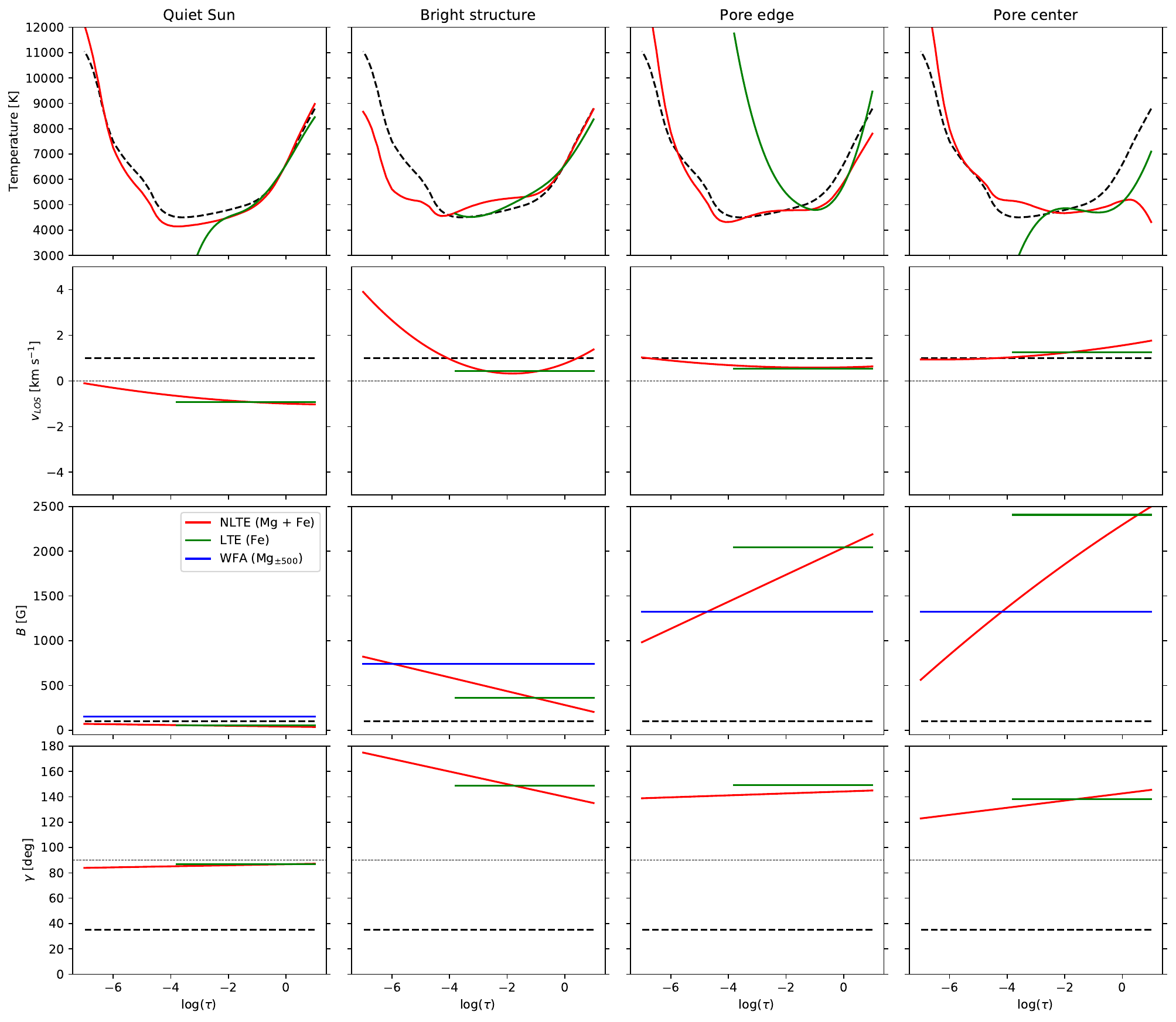} 
      \caption{Atmospheric stratifications obtained with the different methods for the Stokes profiles shown in Fig. \ref{fig:fits_SP}:
      the initial FALC model (dashed black lines) and the results from the two-line NLTE inversions (red lines), from the WFA using the spectral region $\pm500$ m\AA\ around the Mg line core (blue), and from the LTE inversion of the Fe line (green). Panels display, from top to bottom, the temperature in K, the LOS velocity in km s$^{-1}$, the magnetic field strength in G, and the magnetic field inclination in degree. 
              }
         \label{fig:atms_SP}
   \end{figure*}

 \begin{figure*}[hbtp!]
   \centering 
    
     \includegraphics[width=\hsize]{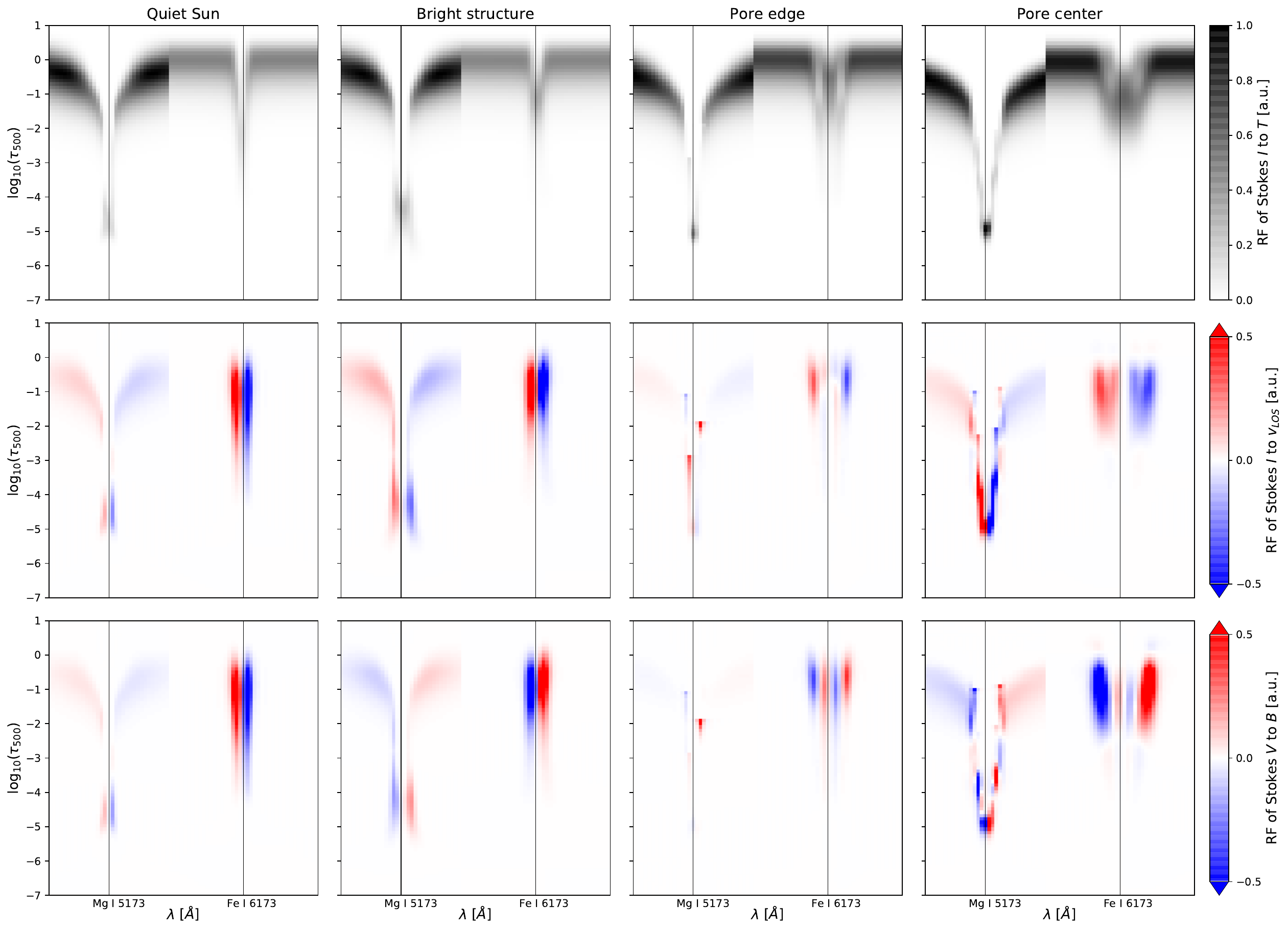} 
      \caption{NLTE RFs computed for the different sets of Stokes profiles and resulting atmospheres shown in Figs. \ref{fig:fits_SP} and \ref{fig:atms_SP}, respectively. From top to bottom, the maps show the RF of Stokes $I$ to changes in temperature, of Stokes $I$ to changes in LOS velocity, and of Stokes $V$ to changes in the magnetic field strength. All maps are normalized to their maximum value.
                  }
         \label{fig:RFs_SP}
   \end{figure*}

Our tests confirm that NLTE inversions perform better when incorporating as much spectral and atomic information as possible. However, the computational time is significantly reduced when  simplified atomic models are used and  fewer spectral lines and/or wavelength points are considered. This reduction in computation time becomes especially important when performing NLTE inversions on full Stokes spectra over a large FOV or in long time series. Therefore, the best inversion strategy depends on the specific case and the main objectives of the study.

\subsection{NLTE inversions: Large FOV}

As mentioned before, conducting NLTE inversions is computationally expensive, especially when dealing with 
full Stokes spectropolarimetric observations that cover a large FOV. Therefore, we performed a simple inversion for our two-line  dataset using only five cycles. The inversion employed a one-component FALC initial model and assigned different weights to the Stokes parameters based on their signal-to-noise ratio: weights of 10 for Stokes $I$ and $V$, and weights of 1 for Stokes $Q$ and $U$. 

Considering the results from Sect. 3.4, we used the 13-level atomic model and applied OF, inverting all cycles with a NLTE threshold of 0 to treat the Mg atom fully in NLTE (the Fe atom was treated in LTE).
The temperature was allowed to vary over a maximum of six nodes, the microturbulence over five nodes, and we used up to three nodes for the LOS velocity and the magnetic field parameters.

\section{Results}

\subsection{NLTE inversions of Mg I b$_2$ 5173 \AA\ and Fe I 6173 {\AA}}
Figure \ref{fig:NLTE_FOV} presents the results of the simultaneous NLTE inversion of the Mg I 5173 and Fe I 6173 lines across the entire FOV, at five different optical depths, for the temperature, LOS velocity, unsigned longitudinal magnetic field, and magnetic field inclination.
The typical granulation pattern of the photosphere is well retrieved as displayed by the temperature and LOS velocity maps of the surface layers, at $\log(\tau)=0$. Higher up, at $\log(\tau)=-1$, there is a mean temperature decrease of nearly 1300 K and the inverse granulation pattern is observed, with granules appearing about 300 K cooler than the inter-granules. The plage around pores is nearly 700 K hotter than the surroundings at this height.
The mean temperature continues to decrease with height, while the temperature contrast between the different structures becomes more pronounced. In the velocity maps, the granular flows weaken progressively at higher layers. However, at $\log(\tau)=-4$, stronger flows forming fibril-like structures between the opposite-polarity pores are inferred. 
The maps of the magnetic field parameters, both $|B_{LOS}|$ and $\gamma$, show that the areas covered by the magnetic structures appear more compacted in the photosphere and gradually become broader with height. 
This behavior resembles the chromospheric canopy of larger magnetic structures, such as sunspots \citep[e.g.,][]{Joshi2017b}, but here it occurs at lower heights, meaning that the field rapidly bends over and becomes more horizontal in the mid-to-high photosphere.  
The strongest longitudinal fields are concentrated within the core of the pores, where they reach strengths on the order of 2 kG in the photosphere, and gradually become weaker with height.
Strong fields are also observed in the plage areas as well as in some isolated bright points, and can reach strengths on the order of 1 kG. 
The results from our inversions are qualitatively satisfactory and provide, on large scales, a reliable  stratification of the magnetic and thermodynamic parameters around the temperature minimum region. 

Figure \ref{fig:chi2} shows the map of the merit function, $\chi^2$, for the best fits of the inversions, computed as follows: 


\begin{equation}
\chi^2 \equiv \frac{1}{\nu}\sum_{k=1}^{4}\sum_{i=1}^{M}{\frac{[I_{k}^{obs}(\lambda_i)-I_{k}^{syn}]^{2}}{\sigma_{k,i}^2}}
\label{eq_chi}
,\end{equation}

\noindent which is the sum of the squared differences between observed and synthetic profiles, $I^{obs}$ and $I^{syn}$, respectively, weighted by the noise of the observations, $\sigma$ -- which is Stokes- and line-dependent -- and the number of degrees of freedom $\nu$. 
In Eq. \ref{eq_chi}, the index $k=1,...,4$ samples the four Stokes profiles, while the index $i=1,...,M$ samples the wavelengths at which the line has been measured.
 The map indicates generally good quality fits in most of the FOV, with an average $\chi^2<2.5$ in the quiet Sun. However, in approximately $0.06\%$ of the map, the code could not converge to a good solution and no synthetic Stokes profiles were generated (red pixels in Fig.\ \ref{fig:chi2}). 
 In addition, the inversion yields poorly fitted profiles ($\chi^2>10$) in different places but predominantly in the center of the pores. This mostly happens when the observed profiles display uncommon features, such as very large asymmetries, very shallow profiles, emission in the wings, extra lobes in Stokes $V$, among others, likely due to the relative simplicity of our inversion configuration. 
  Nonetheless, the quality of the fits is good in most of the FOV. In Fig. \ref{fig:fits_SP} we present some examples of typical NLTE fits to sample Stokes profiles emerging from the quiet Sun, a bright structure, a pore's edge, and the magnetic core of a pore. The figure also displays the fits obtained from the LTE inversions (for the Fe I line only) and from the WFA using the spectral window of $\pm500$ m\AA\ in the Mg I b$_2$ line (for Stokes $V$  only). All three methods are able to closely reproduce the observed profiles and provide comparable fits. Nonetheless, the NLTE inversions generally provide better fits than the WFA for the Mg Stokes $V$ profile, particularly in the wing wavelengths.

We recall that we kept the NLTE inversions as simple as possible in order to speed up the computations, so we used a few cycles, a small number of nodes in the physical parameters, a single-component atmospheric model, and the same initial setup to fit an enormous variety of shapes and different spectropolarimetric features in two spectral lines simultaneously. 
Alternative ways to perform NLTE inversions to improve the quality of the fits is out of the scope of this paper but will be addressed in detail in an upcoming second paper.

 The atmospheric stratifications resulting from these fits are shown in Fig. \ref{fig:atms_SP}. In general, the temperature variation with height in the LTE inversions aligns well with the NLTE results in the photospheric layers, but there are significant discrepancies above $\log(\tau)\sim-2$, especially in the pores. However, the LOS velocity, magnetic field strength, and magnetic field inclination show good agreement between the LTE and NLTE inversions in the lower layers of the atmosphere.
 Additionally, the magnetic field strength derived using the WFA in the Mg $\pm500$ m\AA\ spectral window matches that retrieved from the NLTE inversions in the higher layers, so that both the WFA and the LTE inversions qualitatively align with  
   the estimated  magnetic field gradients obtained from the NLTE inversions.

 Figure \ref{fig:RFs_SP} shows the NLTE response functions (RFs) computed for the Stokes profiles and the atmospheric models presented in Figs. \ref{fig:fits_SP} and \ref{fig:atms_SP}, respectively. The figure shows that the sensitivity of the lines is different for each set of profiles, both in the core and the wings of the Mg I and Fe I lines. The height of maximum response also changes with the physical parameter. 
As a result,  it is not feasible to select a single representative optical depth based on RFs in a specific spectral region that works for all solar structures and for all the physical parameters.
However, a representative optical depth is necessary for a quantitative comparison between methods.
 Therefore, in the following subsections, we compare the results from the different methods at the heights where the cross-correlations between them are highest over the full FOV.

 \begin{figure*}[hbtp!]
   \centering
  
    \includegraphics[width=0.55\hsize]{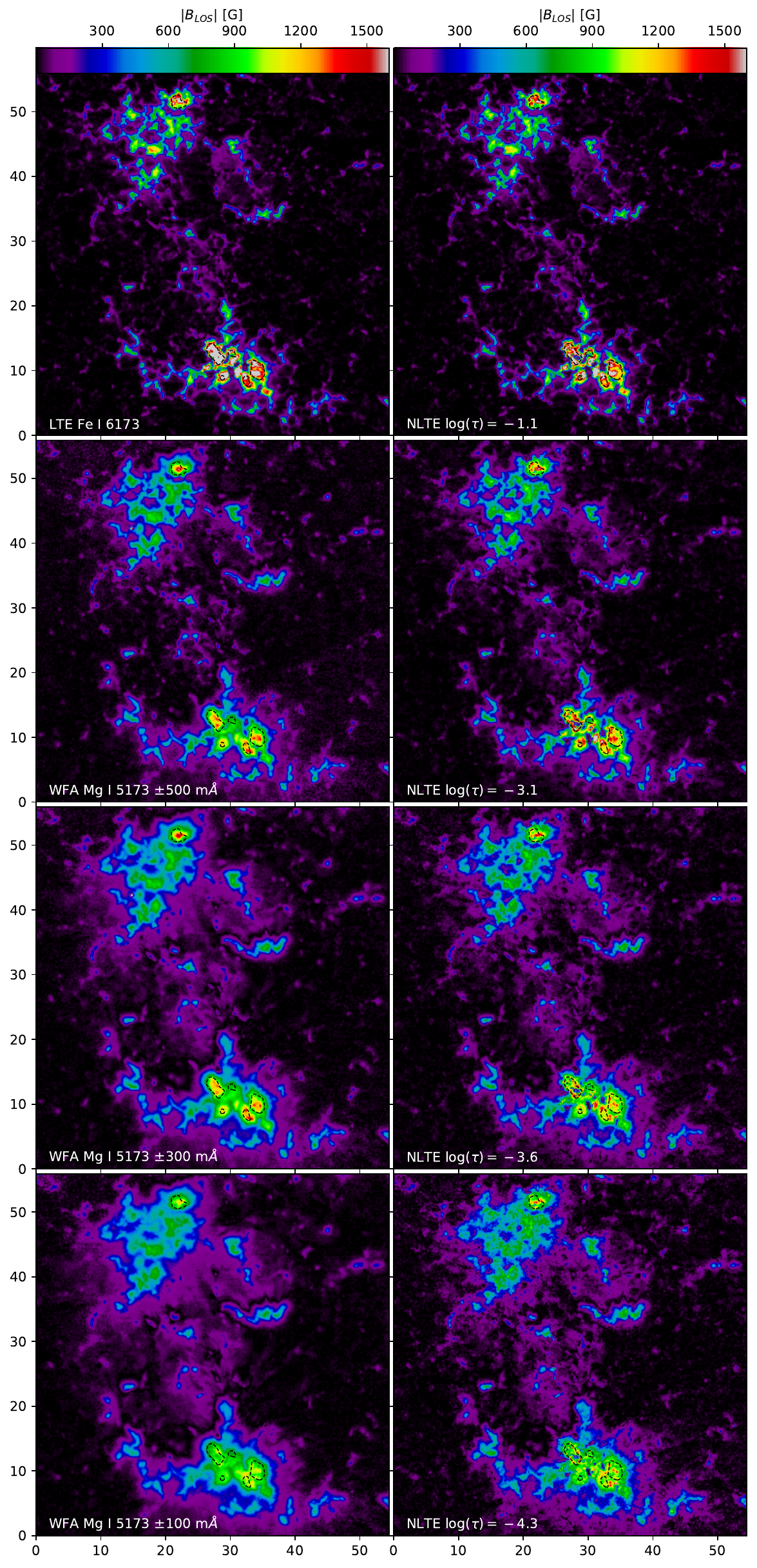} 
       \includegraphics[width=0.37\hsize]{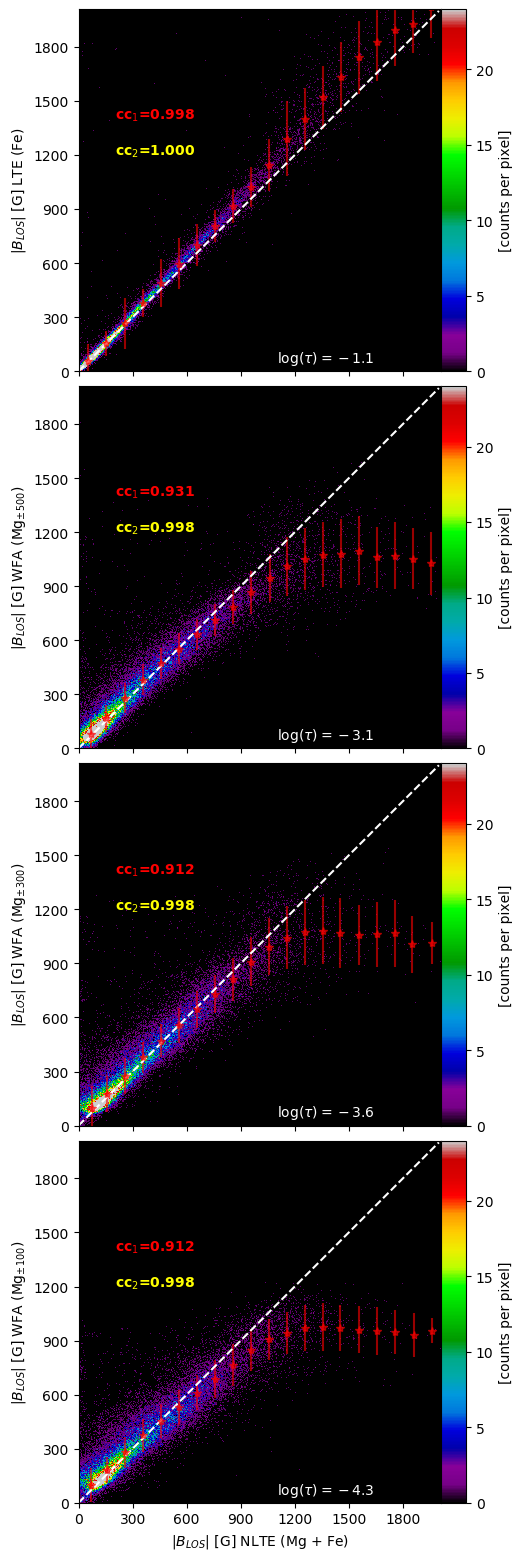}%
      \caption{Height variation of the longitudinal magnetic field, comparing results from the different methods. In the left column we show the traditional methods (from top to bottom): LTE inversion of the Fe I 6173 \AA\ line, and  WFA applied to the Mg I 5173 \AA\ line within the spectral windows of $\pm500$ m{\AA}, $\pm300$ m{\AA}, and $\pm100$ m{\AA}. In the center we show NLTE inversion at different optical depths, selected based on the maximum cross-correlation with the corresponding left maps, and in the right scatter plots comparing the results from the different methods, excluding $|B_{LOS}|$ values obtained from Stokes $V$ profiles with signals below the 5$\sigma$ level.      
     Black contours on the left and central maps show the position of the pores.  The dashed white lines in the scatter plots represent identity lines,  and the red markers indicate the average values of the distribution in x-bins of 100 G. The red vertical lines depict the standard deviation along the y-axis for each 100 G bin on the x-axis. Two correlation coefficients are provided in each scatter plot: $cc_1$ for the entire distribution and $cc_2$ for $|B_{LOS}|<1000$ G.}
         \label{fig:3a}
   \end{figure*}

 \begin{figure*}[hbtp!]
   \centering
    
     \includegraphics[width=0.59\hsize]{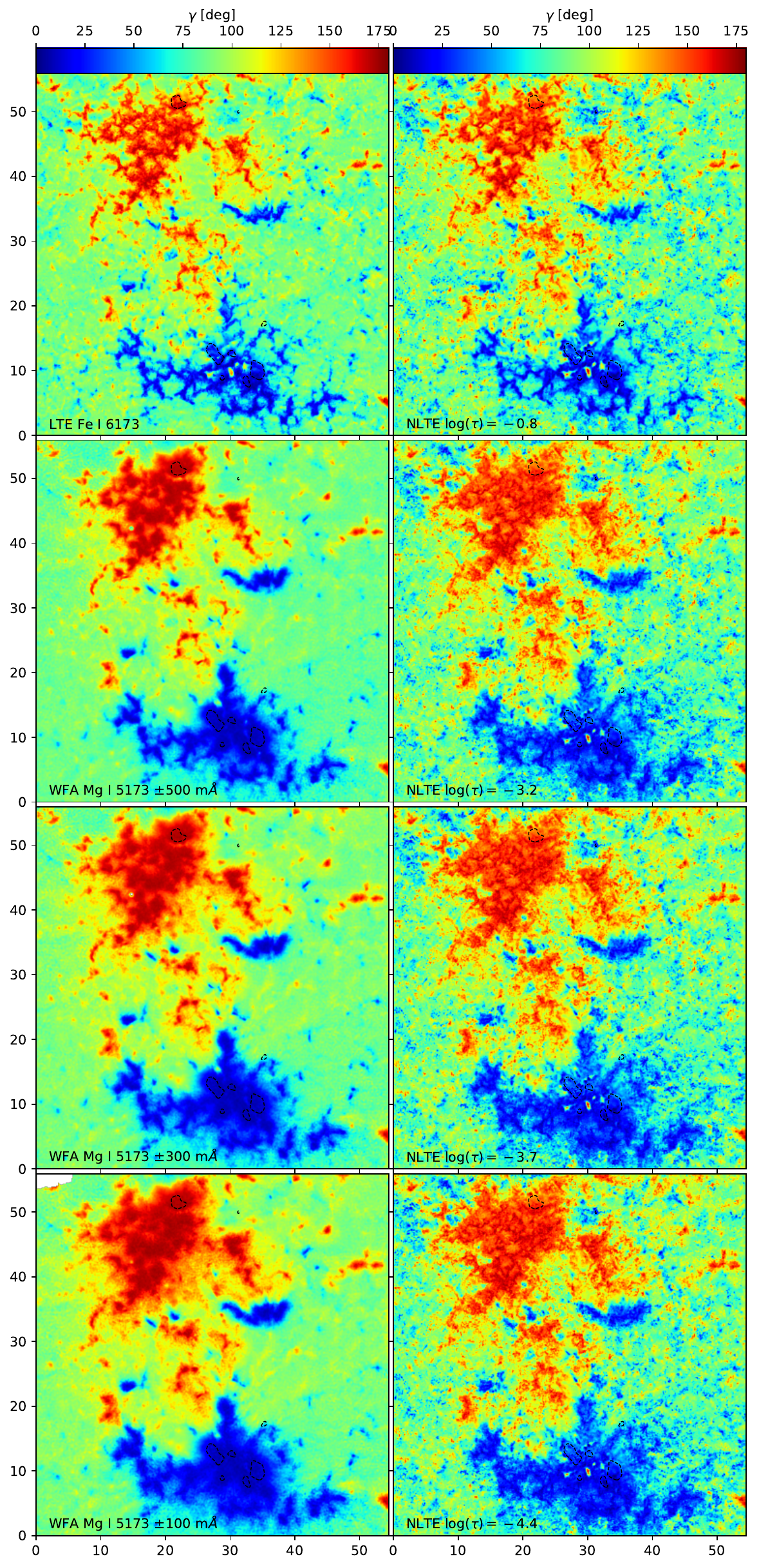} 
    \includegraphics[width=0.39\hsize]{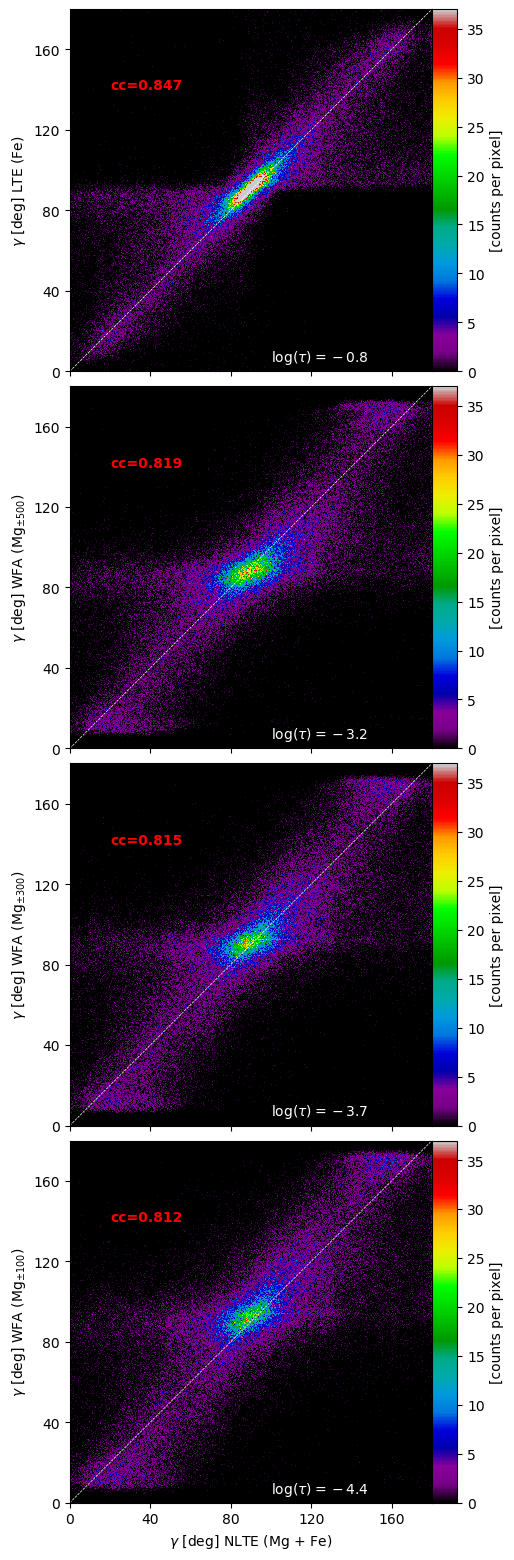} 
      \caption{
      Height variation of the magnetic field inclination, comparing results from the LTE and WFA methods with those from the NLTE inversions. The format is similar to that of Fig.\ \ref{fig:3a}.}
         \label{fig:comp_INC}
   \end{figure*}

\subsection{Magnetic field stratification: LTE and WFA versus NLTE}

Figures \ref{fig:3a} and \ref{fig:comp_INC} 
 compare the results for the unsigned LOS magnetic field and magnetic field inclination, respectively, obtained from the different methods.
In the first column of each figure, we present the results from the traditional methods: LTE inversion of the Fe I 6173 \AA\ line and WFA applied to the Mg I 5173 \AA\ line using progressively smaller spectral windows around the line core, roughly ordered by increasing formation height. 
The second column displays the results from the NLTE inversions at the optical depths that are best comparable with the results shown in the first column, and correspond to those displaying the largest correlation coefficients, $cc$.
These optical depths are $\log(\tau)=-1.1$, $-3.1$, $-3.6$, and $-4.3$ for the $|B_{LOS}|$ maps, and  $\log(\tau)=-0.8$, $-3.2$, $-3.7$, and $-4.4$ for the $\gamma$ maps.

 The $|B_{LOS}|$ maps in Fig.\ \ref{fig:3a} show good qualitative agreement between methods, 
with the magnetic structures appearing remarkably similar in the first and second column maps. Both traditional and NLTE approaches consistently reveal the expansion and opening of the magnetic field with height around the magnetic structures, as well as the weakening of the magnetic field away from the core of pores, plage, and bright magnetic points.
We refer to the regions where $|B_{LOS}|>200$ G as ``canopy areas''. The growth with height of the canopy areas with respect to the LTE maps is of 147$\%$, 182$\%$, and 189$\%$ in the maps from the WFA in the $\pm500$ m{\AA}, $\pm300$ m{\AA}, and $\pm100$ m{\AA} spectral windows, respectively; while it is of 143$\%$, 180$\%$, and 187$\%$ in the NLTE maps at $\log(\tau)=-3.1$, $-3.6$, and $-4.3$, respectively, with respect to $\log(\tau)=-1.1$.

The scatter plots in the third column provide a quantitative comparison between methods.
In the scatter plots of Figs.\ \ref{fig:3a} and \ref{fig:comp_INC}, we compare only the $|B_{LOS}|$ and $\gamma$ values, respectively, obtained in pixels where the Stokes $V$ profiles show signals above five times the noise level.
The correlation coefficients for $|B_{LOS}|$ indicate good  agreement at all four heights, particularly for weaker magnetic fields, although this correlation slightly decreases  with increasing height --- likely due to the small number of wavelengths sampled within the line core, which affects the performance of both methods.
Additionally,  the scatter plots show that the one-to-one correspondence between methods decreases notably for longitudinal fields stronger than 1000 G. 
This trend is evident in the mean values of the 100 G bins in Fig.\ \ref{fig:3a}, which progressively deviate from the identity line at stronger fields. Likewise, 
the standard deviation, $std$, of the values grows with field strength, reflecting increasing differences between methods at higher $|B_{LOS}|$ values.

On the one hand, this discrepancy  manifests as the LTE inversions yielding stronger $|B_{LOS}|$ values than those from the NLTE inversions at $\log(\tau)=-1.1$.
Specifically, for $|B_{LOS}| \approx 1000$ G, the average difference between methods is on the order of 70 G, with a $std$ of nearly 110 G. This discrepancy peaks at $|B_{LOS}| \approx 1600$ G, where the average difference reaches nearly 180 G, and the $std$ approaches 200 G. 
 These differences arise mainly because the LTE inversions assume a constant magnetic field, while the NLTE inversions account for magnetic field gradients with height, leading to greater discrepancies in regions with strong gradients. 

On the other hand, the WFA applied to the three different spectral windows in the Mg line generally infers weaker $|B_{LOS}|$ values in strong-field regions compared to the NLTE inversions at $\log(\tau)=-3.1$, $-3.6$, and $-4.3$, respectively.  The WFA appears to saturate around an average $|B_{LOS}|$ of 1100 G for the  $\pm 500$ and  $\pm 300$ m\AA\ spectral windows, and around 1000 G for the  $\pm 100$ m\AA\ spectral window. 
These discrepancies are partly due to the limitations of the WFA, which is only valid for weak magnetic fields and assumes the absence of gradients along the LOS in the magnetic field and in the plasma velocities --- conditions that are likely not met in the bright magnetic regions and the pores, as suggested by the NLTE inversion results. 
Another contributing factor to this discrepancy could be the poorer quality of the NLTE  fits in the core of pores, as shown by the $\chi^2$ map in Fig.\ \ref{fig:chi2}. At such places, the intensity profiles appear very shallow or even show strong apparent wing emissions --- characteristics that are not properly reproduced due to the relative simplicity of the NLTE inversions. Thus, it is possible that  $|B_{LOS}|$ is  overestimated by the NLTE inversion in those cases.

Except for the core of the pores, the bulk of the distribution from the traditional methods show both qualitative and quantitative consistency with the NLTE inversions ($cc_2$ values in Fig.\ \ref{fig:3a}), confirming the validity of the WFA as a viable alternative  for the rapid inference of the height stratification of the longitudinal magnetic field within the Mg line formation region, up to strengths on the order of 1000-1100 G, depending on the spectral window around the line core wavelength.

 \begin{figure*}[hbtp!]
   \centering
   
    \includegraphics[width=0.59\hsize]{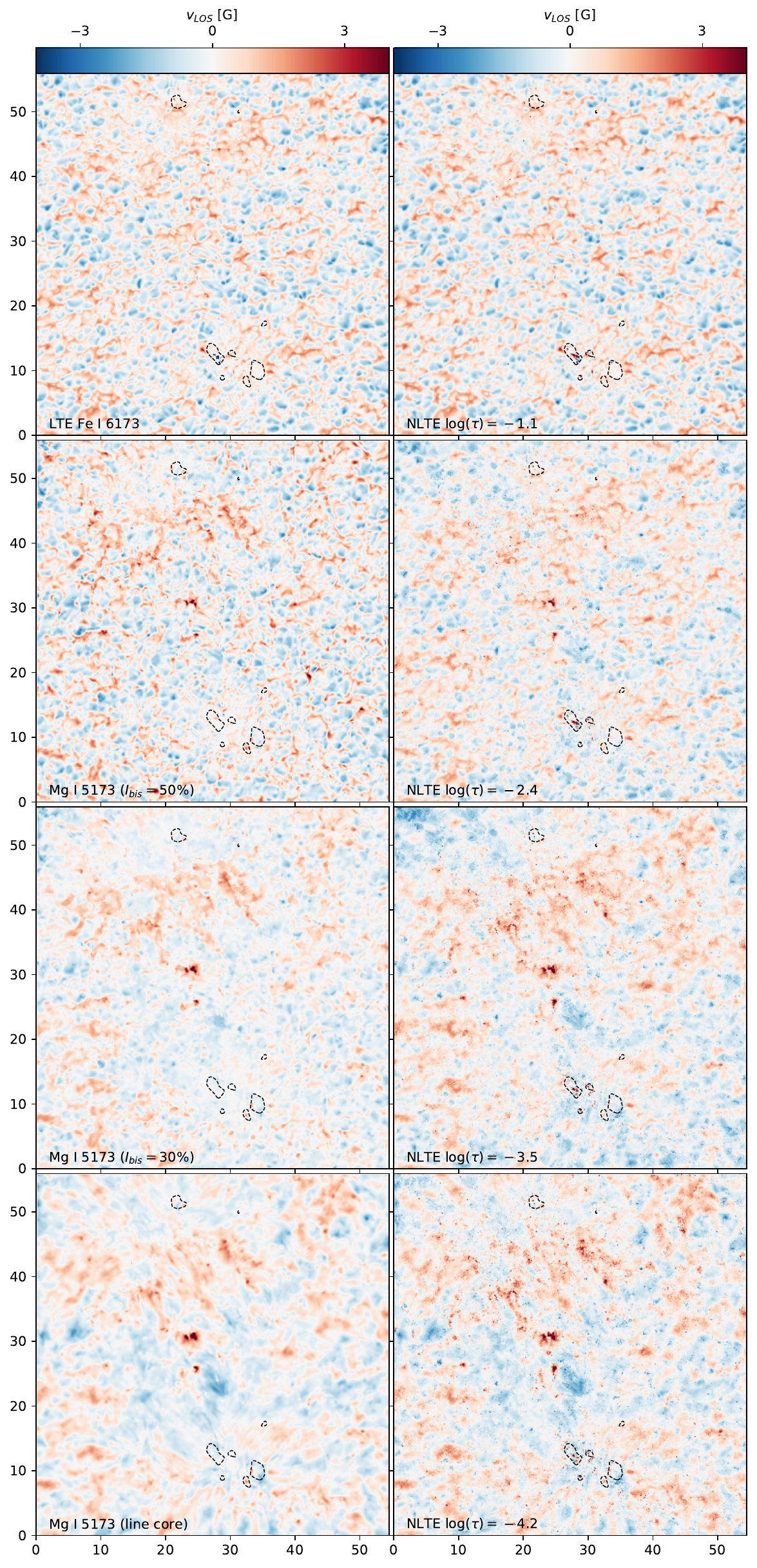}
        \includegraphics[width=0.39\hsize]{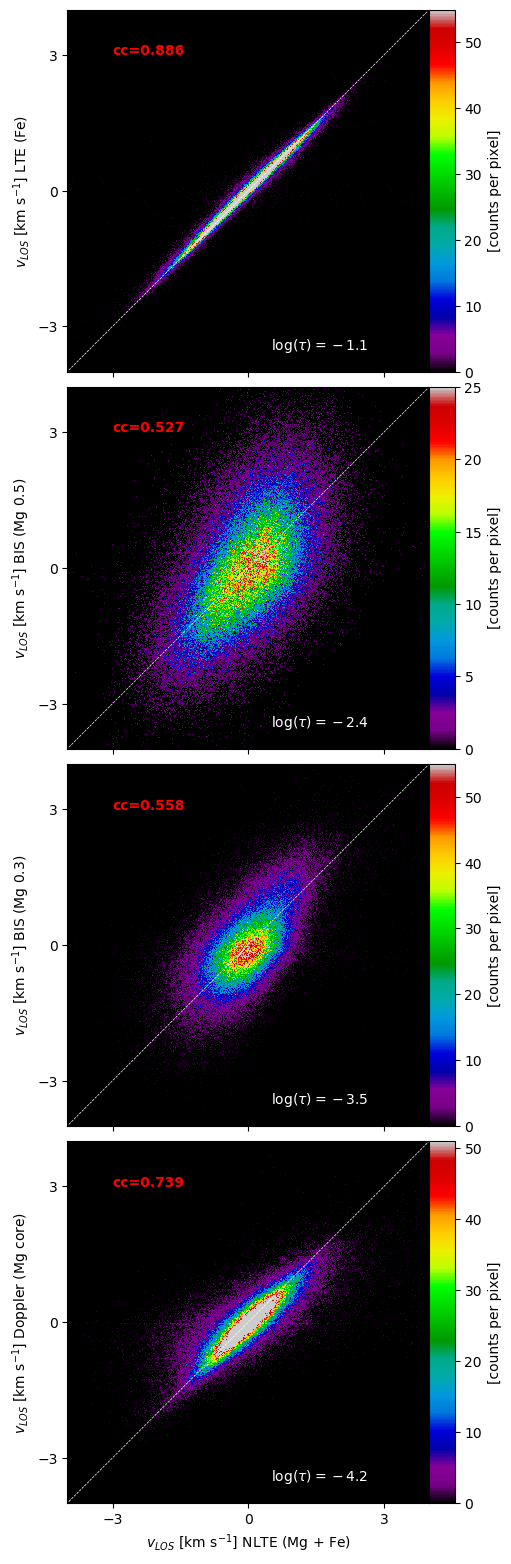}
      \caption{Height variation of the LOS velocity, comparing results from the different methods. On the left we show traditional inferences (from top to bottom):  LTE inversions of the Fe I 6173 \AA\ line,  bisector velocity of Mg at $50\%$, bisector velocity of Mg at $30\%$, and Doppler velocity of the Mg line core.\ In the center we show NLTE inversion results at different optical depths, selected based on the maximum cross-correlation with the traditional inferences, and on the right scatter plots comparing the results from the different methods. The format is similar to that of Fig.\ \ref{fig:3a}.}
         \label{fig:comp_VLOS}
   \end{figure*}

The comparison of the magnetic field inclination in Fig.\ \ref{fig:comp_INC} also shows qualitative similarities between methods, particularly in the broadening of the magnetic regions with height, and shows consistency regarding the main polarity of the different magnetic structures across the FOV. 
However, notable discrepancies emerge mainly in the quiet Sun areas, where the NLTE inversions yield a broader range of inclinations, giving the maps a noisier appearance. In contrast, the LTE inversion and the WFA predominantly infer horizontal fields in these regions.
Assessing the reliability of each method in this context is challenging, as the weak linear polarization signals from both lines in the quiet Sun are essentially at the noise level. 
Under these conditions, all the methods are expected to result in nearly horizontal fields that are not real but a consequence of the noise \citep[e.g.,][]{Lites2017}.
This limitation substantially affects the performance of all the WFA, LTE, and NLTE inversions to accurately retrieve the transverse component of the magnetic field in the quiet Sun.
 Nonetheless, the scatter plots in Fig.\ \ref{fig:comp_INC} show that the correlation between methods is strong for the magnetic structures, as indicated by the $cc$ values, provided their linear polarization signals are larger than the noise level.

\subsection{LOS velocity: LTE and LB versus NLTE}
Figure \ref{fig:comp_VLOS} displays LOS velocity maps derived from different methods. In the left column, we present traditional inferences of $v_{LOS}$ for different heights in the atmosphere: LTE inversions of the Fe line, LB at $50\%$ of the Mg line, LB at $30\%$ of the Mg line, and the Doppler velocity derived from the Mg line core. As with the magnetic field comparison,   these velocity maps are compared with those resulting from the two-line NLTE inversions at the optical depths where they are better correlated, as shown in the second column of the figure. The selected optical depths are $\log(\tau)=$-1.1, -2.4, -3.5, and -4.2, respectively.

The scatter plots in the third column show a strong correlation in the photosphere,  between the LTE inversion of the Fe I line and the NLTE inversions at $\log(\tau)=$-1.1, as well as in the chromosphere, between the Mg I line core Doppler velocities and the NLTE inversion at $\log(\tau)=$-4.2.
However, the correlations between the NLTE velocities and bisector velocities, while qualitatively consistent, are less quantitatively robust.

Although LBs can generally be associated with the LOS velocities at different heights — with intensity levels closer to the continuum relating to motions in photospheric layers and levels nearer to the line core sensing LOS velocities at higher layers — when applied to the spectral lines typically used for photospheric diagnostics, in this case the bisector method is not sufficiently good for the Mg I b$_2$ line.  
This limitation likely arises because the Mg I b$_2$ line is primarily chromospheric in nature and exhibits significant opacity effects. High chromospheric dynamics (e.g., umbral flashes, Ellerman bombs) often cause asymmetric line profile emissions. Additionally, the presence of various photospheric blends within the Mg I b$_2$ line may introduce further distortions in the line's profile. These combined effects alter the line shape and disrupt any linear relationship between the LBs and the LOS velocities.

\section{Discussion and conclusions}

Testing the validity of methods that are simpler and faster than NLTE inversions to extract physical information about the solar atmosphere from the Mg I b$_2$ line at 5173 $\r{A}$ and other chromospheric spectral lines has become increasingly relevant with the advent of large datasets from new-generation solar observatories like SUNRISE III \citep{Lagg2024}, DKIST \citep[e.g.,][]{Rimmele2019}, EST \citep[][]{Quintero2022}, and others.
Some of these facilities are now providing high-resolution solar observations in the Mg I b$_2$ line, which probes the temperature minimum region --- a  region of the solar atmosphere whose magnetism is still relatively unexplored  due to the limited number of spectral lines with sufficient magnetic sensitivity that form at those heights \citep[e.g.,][]{Quintero2018}.

In these instances, a fast yet reliable estimation of the magnetic field and the LOS velocity height stratification can be valuable 
for quick inspection purposes. 
In the literature, different methods have been tested on the Mg I b$_2$ line to obtain a fast computation of the magnetic field using synthetic data \citep[e.g.,][]{Morosin2020, Dorantes2022, Vukadinovic2022}.
In this work we evaluated the capabilities of traditional  methods to infer the height stratification of the magnetic field and of the LOS velocities from the Mg I b$_2$ line, using high-resolution spectropolarimetric observations in an extended FOV. 
Specifically, we inferred the height variation of the longitudinal component of the magnetic field and of the magnetic field inclination by combining an LTE inversion of the Fe I line at 6173 \AA\ with the WFA in three spectral windows within the Mg I b$_2$ line. 
Likewise, the stratification with height of the LOS velocities were computed by combining the LTE inversion of the Fe I line with bisector velocities at different intensity levels of the Mg I b$_2$ line and with Doppler velocity maps of the Mg line core. 

In addition, we conducted a full NLTE inversion of the  Mg I b$_2$ line at 5173 \AA\ simultaneously with the photospheric Fe I line at 6173 {\AA}, using the DeSIRe inversion code \citep{Ruiz2022}. This is the first time that high-resolution spectropolarimetric observations of the Mg I b$_2$ line have been inverted under a NLTE scheme in a large FOV using the full Stokes vector. 
Despite the relative simplicity of the inversions and the long computation times that were required, the generally good quality of the fits 
lends confidence to the resulting stratifications, establishing a reliable basis for comparison with traditional inference methods.

We find that the WFA applied to progressively narrower spectral windows within the Mg I b$_2$ line provides not only a consistent overview of the magnetic field variation with height in the low chromosphere, but also reliable quantitative information about the chromospheric longitudinal field component at weak and moderate strengths. 
The LTE inversions of the Fe I 6173 \AA\ line and the WFA results for the Mg I b$_2$ 5173 \AA\ line align well with the NLTE inversions in capturing the expansion of the canopy areas with height, specifically at $\log(\tau)=-1.1$, $-3.1$, $-3.6$, and $-4.3$.  
However, attention is required for strong magnetic field concentrations, such as pores and umbral regions, since in those regions the WFA  
does not seem to be valid, resulting in saturation of the field strength values, even if magnetic fields are typically weaker in the chromosphere than in the photosphere \citep[e.g.,][]{Joshi2016}.
In addition,  chromospheric spectral lines typically have a larger Doppler width than photospheric lines, a characteristic that makes the WFA generally valid for chromospheric lines in the Zeeman regime \citep[e.g.,][]{delaCruz2013, Centeno2018, Morosin2020}.

Recently, \citet{Vukadinovic2022} investigated the applicability of the WFA using synthetic spectra of the Mg I b$_2$ line and found that this method can reliably retrieve longitudinal fields up to 1400 G with an error lower than 5$\%$ (about 70 G) in atmospheres with no velocity gradients. 
In our analysis, where velocity gradients are present, we find that the $|B_{LOS}|$ values estimated from the WFA in the Mg I b$_2$ spectral windows that probe layers in the upper photosphere and lower chromosphere
 show  discrepancies of a few hundred gauss when compared with the NLTE results in the core of the pores, being, on average, about 300 G weaker when NLTE inversions yield field strengths of 1400 G. 
 These discrepancies suggest larger errors in the WFA results compared to those estimated by \citet{Vukadinovic2022}. Nonetheless, such discrepancies could also partly originate from the NLTE inversions, which did not achieve particularly good fits in the pore regions, especially in their magnetic cores.

The LTE inversions and the WFA were also used to infer the magnetic field inclination at different heights in the atmosphere.
The results are consistent and show strong correlations with the NLTE inversions regarding the main polarity of the magnetic structures and the expansion of the magnetic canopies around them --- specifically at $\log(\tau)=-0.8$, $-3.2$, $-3.7$, and $-4.4$.
However, we find weaker correlations in the quiet Sun regions, where the traditional methods predominantly infer horizontal fields, unlike the NLTE inversions, which retrieve a mixture of polarities in these areas.
 Nevertheless, without sufficiently strong linear polarization signals above the noise level, the WFA, LTE, and NLTE inversions cannot properly estimate the transverse magnetic field, nor the magnetic field inclination. 
Previous studies have suggested that the WFA may not be suitable for estimating the transverse magnetic field in the Mg I b$_2$ line, given that this is a strong absorption line, whilst the nonlinear relationship between linear polarization and the transverse field component makes the WFA of the transverse field more appropriate for  weak absorption lines \citep{Bai2013}. 
Nonetheless, our results indicate that in magnetic regions where the linear polarization signals exceed the noise level, the WFA can provide both qualitatively and quantitatively reliable estimates of the magnetic field inclination from the Mg I b$_2$ line.
 
 We also tested the LB method \citep[][]{Maltby1964} and line-core Doppler shifts in the Mg line to estimate the variation with height of the LOS velocities.
 Our findings indicate that the LBs might not accurately retrieve LOS velocities in the Mg I b$_2$ line. This may stem from 
significant opacity effects and/or line distortions that could be produced by a large variety of phenomena other than  plasma motions, such as wave propagation, shocks, flashes, heating, among others which are common in the highly dynamic chromospheric layers and might affect the shape of the intensity profile  \citep[e.g.,][]{Uitenbroek2006}. 
Furthermore, photospheric blends present in the Mg I b$_2$ line can 
frequently distort the profile further, complicating a straightforward velocity interpretation. 
As a result, the LB method appears unreliable for rapid and precise  LOS velocity estimates from the Mg I b$_2$ line. However, information derived from the Fe I line and the Doppler shifts of the Mg I b$_2$ line core demonstrates good agreement with NLTE inversions at $\log(\tau)=-1.1$ and $-4.2$, respectively. These methods are therefore still viable alternatives for inferring LOS velocity gradients.

Finally, our inversion tests using the FTS atlas profiles have shown that NLTE inversions yield better fits and more realistic atmospheres when more detailed information is included. Specifically, for the Mg I b$_2$ line, the inversions perform optimally when a more complex atomic model is used along with sufficient photospheric spectral information (from spectral lines or blends), an OF correction, larger spectral windows, and adequate wavelength sampling. All these elements must be carefully optimized based on the specific science goals and the computational resources available.

\begin{acknowledgements}
      We thank Basilio Ruiz Cobo,  Carlos Quintero Noda, Han Uitenbroek, Ricardo Gafeira, and Jaime de la Cruz Rodr\'iguez, for valuable discussions and useful assistance on the NLTE inversions. A.S.T. has been funded by Consejer\'ia de Transformaci\'on Econ\'omica, Industria, Conocimiento y Universidades from Junta de Andaluc\'ia through grant POSTDOC-21-00832. This work was supported by the Spanish Ministry of Science and Innovation through project PID2021-125325OB-C51. The authors acknowledge financial support from the Severo Ochoa grant CEX2021-001131-S funded by MICIU/AEI/10.13039/501100011033.  The Swedish 1 m Solar Telescope is operated on the island of La Palma by the Institute for Solar Physics of Stockholm University in the Spanish Observatory del Roque de los Muchachos of the Instituto de Astrofísica de Canarias. The Institute for Solar Physics is supported by a grant for research infrastructures of national importance from the Swedish Research Council (registration number 2017-00625).

\end{acknowledgements}

%
 \bibliographystyle{aa}
\bibliography{Mgclass.bib}







   
  



\end{document}